\documentclass[acmsmall,screen,nonacm]{acmart}

\usepackage{amssymb}
\usepackage{amsmath}
\usepackage{amsthm}
\usepackage[T1]{fontenc}
\usepackage{mathpartir}
\usepackage{stmaryrd}
\usepackage{trimclip}
\usepackage{xcolor}
\usepackage{listings}
\usepackage[skins]{tcolorbox}
\usepackage{tikz-cd}
\usetikzlibrary{decorations.pathmorphing} \usepackage{placeins}
\usepackage{wrapfig}

\definecolor{extcolor}{rgb}{0.75, 0.30, 0.00}  \definecolor{modcolor}{rgb}{0.00, 0.35, 0.75}  \newcommand{\kw}[1]{\mathbf{#1}}               \newcommand{\ty}[1]{\mathsf{#1}}               \newcommand{\kwe}[1]{\textcolor{extcolor}{\kw{#1}}}
\newcommand{\kwm}[1]{\textcolor{modcolor}{\kw{#1}}}
\newcommand{\exte}[1]{\textcolor{extcolor}{#1}}
\newcommand{\modl}[1]{\textcolor{modcolor}{#1}}

\newcommand{\sce}{\lambda_{\textsf{SCE}}}
\newcommand{\lame}{\lambda_E}
\newcommand{\unit}{\varepsilon}

\newcommand{\Int}{\ty{Int}}

\newcommand{\Intf}{\kwm{sig}}

\newcommand{\marr}{\mathrel{\modl{\to_{m}}}}

\newcommand{\query}{\kw{env}}
\newcommand{\elab}[1]{\llbracket #1 \rrbracket}
\colorlet{sourcejudgment}{extcolor}
\definecolor{corejudgment}{RGB}{0,105,75}
\newcommand{\svdash}{\mathrel{\textcolor{sourcejudgment}{\vdash}}}
\newcommand{\cvdash}{\mathrel{\textcolor{corejudgment}{\vdash}}}
\newcommand{\svDash}{\mathrel{\textcolor{sourcejudgment}{\Vdash}}}

\newcommand{\EBig}[3]{#1 \cvdash #2 \Rightarrow #3}
\newcommand{\SBig}[3]{#1 \svdash #2 \Rightarrow #3}
\newcommand{\EDiv}[2]{#1 \cvdash #2 \Uparrow}
\newcommand{\SDiv}[2]{#1 \svdash #2 \Uparrow}
\definecolor{elabbackground}{gray}{0.88}
\newcommand{\cterm}[1]{\mathpalette\ctermbox{#1}}
\newcommand{\ctermbox}[2]{\begingroup\setlength{\fboxsep}{1pt}\mathsurround=0pt
  \colorbox{elabbackground}{$#1\color{black}#2$}\endgroup}

\newenvironment{rulefigure}{\normalsize
  \mprset{sep=1em, vskip=0.2ex}}{\par}
\newenvironment{ruleblock}{\par\addvspace{\smallskipamount}\noindent
  \begin{minipage}{\linewidth}\begin{rulefigure}}{\end{rulefigure}\end{minipage}\par\addvspace{\smallskipamount}}
\newcommand{\ruleheading}[2]{\par\addvspace{0.8ex}\noindent
  \makebox[\linewidth][s]{\fbox{$#1$}\hfill\textit{(#2)}}\par\nobreak\vspace{0.6ex}}
\newsavebox{\formalbox}
\newcommand{\formalrow}[1]{\sbox{\formalbox}{$\displaystyle #1$}\ifdim\wd\formalbox>\linewidth
    \PackageWarning{SCE}{Formal row exceeds linewidth by
      \the\dimexpr\wd\formalbox-\linewidth\relax}\fi
  \makebox[\linewidth][c]{\usebox{\formalbox}}}
\newcommand{\syntaxrow}[1]{\par\addvspace{1ex}\begingroup\normalsize
  \noindent\formalrow{#1}\par
  \endgroup\addvspace{1ex}}
\newcommand{\syntaxcat}[1]{\textit{#1}}
\newcommand{\rulegrouplabel}[1]{\par\addvspace{1.2ex}\noindent
  \makebox[\linewidth][r]{\small\itshape (#1)}\par\nobreak\vspace{0.7ex}\nointerlineskip}
\newcommand{\rulelayout}[2]{\if\relax\detokenize{#1}\relax\else\rulegrouplabel{#1}\fi
  \noindent#2\par\addvspace{0.6ex}}
\newcommand{\rulerow}[2][]{\rulelayout{#1}{\formalrow{#2}}}
\newcommand{\rulepair}[3][]{\rulelayout{#1}{\makebox[\linewidth][s]{\begin{minipage}[b]{0.48\linewidth}\centering
      $\displaystyle #2$
    \end{minipage}\hfill
    \begin{minipage}[b]{0.48\linewidth}\centering
      $\displaystyle #3$
    \end{minipage}}}}
\newcommand{\Elab}[4]{\STyp{#1}{#2}{#3} \rightsquigarrow \cterm{#4}}
\newcommand{\HasType}[3]{#1 \cvdash #2 : #3}
\newcommand{\STyp}[3]{#1 \svdash #2 : #3}
\newcommand{\lookupV}{\mathsf{lookupV}}

\newcommand{\selj}[3]{#1.#2 \rightsquigarrow #3}
\newcommand{\cselj}[3]{#1.#2 \rightsquigarrow #3}
\newcommand{\lookup}{\mathsf{lookup}}
\newcommand{\lin}[3]{#1 : #2 \in #3}
\newcommand{\linc}[3]{#1 : #2 \in #3}
\newcommand{\labels}[1]{\mathsf{label}(#1)}
\newcommand{\lrec}[2]{\{#1 = #2\}}
\newcommand{\rcd}[2]{\{#1 : #2\}}

\newcommand{\letb}[2]{\kwe{let}\; #1 \;\kwe{in}\; #2}
\newcommand{\openm}[2]{\kwe{open}\; #1 \;\kwe{in}\; #2}
\newcommand{\mstruct}[1]{\kwm{struct}\; #1}
\newcommand{\mfunctor}[3]{\kwm{functor}_{#1}(#2)\; #3}

\newcommand{\clos}[3]{\langle #1, \lambda#2.\;#3 \rangle}
\newcommand{\mclos}[3]{\langle #1, \kwm{functor}(#2)\; #3 \rangle}
\newcommand{\fatcomma}{\mathbin{\clipbox*{0 -1ex {\width} .66ex}{$\fatsemi$}}}
\newcommand{\mrg}{\fatcomma}
\newcommand{\nmrg}{\mathbin{\exte{\fatcomma\mkern-2mu\fatcomma}}}
\newcommand{\mlink}[2]{#1 \mathbin{\modl{\bowtie}} #2}
\newcommand{\mapp}[2]{#1 \mathbin{\modl{\bullet}} #2}
\newcommand{\sandbox}{{\modl{\blacksquare}}}
\newcommand{\open}{{\modl{\square}}}
\newcommand{\linkc}{\mathsf{link_c}}
\newcommand{\linkarrow}{\rightsquigarrow_{\mathsf{link}}}
\newcommand{\clinkraw}[5]{#2 : #1 \;\mathbin{\modl{\bowtie}}\; #4 : #3 \;\linkarrow\; #5}
\newcommand{\clink}[5]{\clinkraw{#1}{#2}{#3}{#4}{#5}}
\newcommand{\clinkn}[5]{#2 : #1 \;\mathbin{\modl{\bowtie_{\!*}}}\; #4 : #3 \;\linkarrow\; #5}
\newcommand{\hole}{\square}
\newcommand{\ambm}[2]{\mathsf{amb}_{#1}(#2)}
\newcommand{\SStep}[3]{#1 \svdash #2 \longrightarrow #3}
\newcommand{\SMStep}[3]{#1 \svdash #2 \longrightarrow^{*} #3}

\newcommand{\injl}[2]{\kw{inl}_{#1}\; #2}
\newcommand{\injr}[2]{\kw{inr}_{#1}\; #2}
\newcommand{\scase}[3]{\kw{case}\; #1 \;\kw{of}\; #2 \mid #3}
\newcommand{\fixlam}[3]{\kw{fix}(#1 {\to} #2).\; #3}
\newcommand{\fixclos}[4]{\langle #1, \fixlam{#2}{#3}{#4} \rangle_{\!f}}
\newcommand{\tfold}[2]{\kw{fold}\,[#1]\; #2}
\newcommand{\tunfold}[1]{\kw{unfold}\; #1}
\newcommand{\mlinkn}[2]{#1 \mathbin{\modl{\bowtie_{\!*}}} #2}
\newcommand{\linkok}[2]{#1 \svDash #2}
\newcommand{\selpkg}{\mathsf{selpkg}}
\newcommand{\wire}{\mathsf{wire}}

\newcommand{\rulelabelx}[2]{\hypertarget{rule:#1}{\textsc{#2}}}
\newcommand{\rulerefx}[2]{\mbox{\hyperlink{rule:#1}{\textsc{#2}}}}
\newcommand{\rulelabel}[1]{\rulelabelx{#1}{#1}}
\newcommand{\ruleref}[1]{\rulerefx{#1}{#1}}

\newcommand{\lsto}{\mathrel{\mapstochar\relbar\mapsfromchar}}
\colorlet{lschange}{yellow!45}
\newcommand{\lschg}[1]{{\setlength{\fboxsep}{1pt}\colorbox{lschange}{$#1$}}}
\newcommand{\lsstep}[1]{\text{\sffamily\scriptsize\color{black!60}#1}}

\newcommand{\compactcaption}{\captionsetup{skip=5pt}}
\newcommand{\compactfloatend}{\vspace{-8pt}}
\newcommand{\compactdisplayskips}{\setlength{\abovedisplayskip}{3pt}\setlength{\belowdisplayskip}{3pt}\setlength{\abovedisplayshortskip}{0pt}\setlength{\belowdisplayshortskip}{3pt}}
\newcommand{\compactitems}{\setlength{\itemsep}{0pt}\setlength{\parsep}{0pt}}

\definecolor{codekeyword}{RGB}{38,67,96}
\definecolor{codestring}{RGB}{53,99,77}
\definecolor{codecomment}{gray}{0.40}
\lstdefinelanguage{scelang}{
  morekeywords={let,rec,in,fun,if,then,else,match,with,end,type,of,mu,env,
    fold,unfold,box,open,module,struct,sig,functor,sandbox,link,linkall,import},
  morecomment=[n]{(*}{*)},
  morestring=[b]",
  sensitive=true,
}
\lstdefinestyle{scecode}{
  language=scelang,
  basicstyle=\small\ttfamily\mdseries\upshape\color{black},
  keywordstyle=\bfseries\color{codekeyword},
  commentstyle=\color{codecomment},
  stringstyle=\color{codestring},
  columns=fixed,
  basewidth=0.5em,
  keepspaces=true,
  showspaces=false,
  showstringspaces=false,
  showtabs=false,
  tabsize=2,
  upquote=true,
  aboveskip=6pt plus 1pt minus 1pt,
  belowskip=6pt plus 1pt minus 1pt,
  xleftmargin=1em,
  xrightmargin=1em,
}
\newcommand{\code}[1]{\texttt{\upshape #1}}

\colorlet{sceiground}{modcolor!7}
\colorlet{sceground}{black!4}
\colorlet{laterrule}{black!45}
\lstdefinestyle{srcfile}{style=scecode,
  basicstyle=\footnotesize\ttfamily\mdseries\upshape\color{black},
  xleftmargin=0pt, xrightmargin=0pt, aboveskip=0pt, belowskip=0pt}
\lstdefinestyle{scei}{style=srcfile}
\lstdefinestyle{sce}{style=srcfile}
\newcommand{\srcfile}[1]{{\sffamily\scriptsize\color{black!60}\strut#1}}
\tcbset{srcbase/.style={enhanced, sharp corners, boxrule=0pt,
  left=4pt, right=4pt, top=3pt, bottom=3pt, boxsep=0pt,
  before skip=1pt, after skip=0pt}}
\usetikzlibrary{arrows.meta}
\tikzset{nowline/.style={black!85, line width=0.9pt},
  laterline/.style={laterrule, densely dotted, line width=1.1pt},
  intfflow/.style={modcolor, line width=0.8pt, -{Stealth[length=5pt,width=4pt]}}}
\newtcolorbox{srcnow}[3][]{srcbase, colback=#2, equal height group=#3,
  frame code={\draw[nowline] (frame.south west) rectangle (frame.north east);}, #1}
\newtcolorbox{srclater}[3][]{srcbase, colback=#2, equal height group=#3,
  frame code={\draw[laterline] (frame.south west) rectangle (frame.north east);}, #1}
\newcommand{\srcmark}[1]{\tikz[remember picture]\coordinate (#1) at (0,0.5ex);}
\newcommand{\srclegend}{{\sffamily\scriptsize\color{black!70}\tikz[baseline=-0.6ex]\draw[nowline] (0,0) -- (0.6,0);~available now\qquad
  \tikz[baseline=-0.6ex]\draw[laterline] (0,0) -- (0.6,0);~available later}}

\definecolor{wasmcolor}{RGB}{111,66,193}
\colorlet{wasmground}{wasmcolor!7}
\lstdefinelanguage{wasm}{
  alsoletter={._},
  morekeywords={module,import,export,func,type,param,result,local,global,
    data,elem,declare,ref,null,i32,mut,rec,sub,final,struct,field,array},
  morekeywords=[2]{local.get,local.set,local.tee,global.get,i32.const,
    i32.add,i32.sub,i32.mul,i32.eq,ref.func,ref.cast,struct.new,struct.get,
    array.new_data,call,call_ref,if,else,end},
  morecomment=[l]{;;},
  morecomment=[n]{(;}{;)},
  morestring=[b]",
  sensitive=true,
}
\lstdefinestyle{wasmcode}{
  style=scecode, language=wasm,
  basicstyle=\footnotesize\ttfamily\mdseries\upshape\color{black},
  keywordstyle=\bfseries\color{codekeyword},
  keywordstyle=[2]\color{wasmcolor},
  columns=fullflexible,
  xleftmargin=0pt, xrightmargin=0pt, aboveskip=0pt, belowskip=0pt,
}
\newtcolorbox{wasmsource}{srcbase, colback=wasmground, after skip=6pt,
  frame code={\draw[wasmcolor, line width=0.8pt]
    (frame.north west) -- (frame.south west);}}
\lstdefinelanguage{hostjs}{
  morekeywords={const,new},
  morecomment=[l]{//},
  morecomment=[s]{/*}{*/},
  morestring=[b]",
  sensitive=true,
}
\lstdefinestyle{hostcode}{
  style=scecode, language=hostjs,
  basicstyle=\footnotesize\ttfamily\mdseries\upshape\color{black},
}
\newcommand{\comp}[2]{\mathcal{C}\llbracket #1 \rrbracket_{#2}}
\newcommand{\wi}[1]{\textcolor{wasmcolor}{\texttt{#1}}}   \newcommand{\wty}[1]{\texttt{\$#1}}                        \newcommand{\wfield}[1]{\texttt{#1}}                       \newcommand{\ws}{\hspace{0.85em}}                          \newcommand{\wnote}[1]{\hfill\textcolor{black!50}{\texttt{;\kern-0.18em;}\;#1}} \newlength{\wasmwidth}
\newsavebox{\wasmbox}
\newcommand{\wasmblock}[1]{\begingroup\setlength{\fboxsep}{3pt}\sbox{\wasmbox}{\colorbox{wasmground}{\footnotesize
    \begin{tabular}[t]{@{}p{\wasmwidth}@{}}#1\end{tabular}}}\textcolor{wasmcolor}{\vrule width 1.2pt height \ht\wasmbox depth \dp\wasmbox}\usebox{\wasmbox}\endgroup}
 
\title[Linking as Evaluation: Correct Separate Compilation via First-Class Environments]{Linking as Evaluation}
\subtitle{Correct Separate Compilation via First-Class Environments}

\author{Jam Kabeer Ali Khan}
\email{jamkhan@connect.hku.hk}
\orcid{0009-0007-8505-9667}
\affiliation{\institution{Independent}
\country{Hong Kong}
}

\author{Jinhao Tan}
\email{jinhaot@connect.hku.hk}
\orcid{0009-0005-5779-8409}
\affiliation{\institution{The University of Hong Kong}
 \city{Hong Kong}
 \country{China}
}

\author{Bruno C. d. S. Oliveira}
\email{bruno@cs.hku.hk}
\orcid{0000-0002-1846-7210}
\affiliation{\institution{The University of Hong Kong}
 \city{Hong Kong}
 \country{China}
}

\date{}

\setcopyright{none}

\begin{document}
\begin{abstract}
Separate compilation is central to modular software: components are
compiled independently and linked into an executable. Separate
compilation is correct when a linked program behaves like the program
formed by merging its sources. This property is less routine to prove
than type soundness because it requires a formal account of linking.

We present $\sce$, a simply typed calculus in which linking,
modules, and environments are all \emph{first-class}. Programmers
control module linking through a link operator, which a type-directed
elaboration translates into ordinary terms of a core calculus
of first-class environments. The core calculus's existing evaluation
rules then perform the linking: \emph{linking as evaluation}.
Linking independently compiled modules yields the same core term as 
elaborating their source link. Therefore, compilation and linking commute by construction.
As a result, separate compilation correctness follows directly from the
semantics preservation for both static and dynamic linking.
All meta-theory is mechanized in Lean~4 and we implement a compiler to WebAssembly
in OCaml.

\end{abstract}
\keywords{Separate Compilation, Linking, Semantics Preservation, First-Class Environments, Compiler Correctness, Lean~4 Mechanization}
\maketitle

\section{Introduction}

Type soundness \cite{WrightFelleisen1994,TimanyKrebbersDreyerBirkedal2024}
is a widely accepted sanity criterion for statically typed programming
languages, supported by well-established proof techniques and
mechanization frameworks.
Modern software also relies on separate compilation:
components are developed, typechecked, and compiled independently,
then linked into larger systems
\cite{10.1145/174675.176926,10.1145/178243.178245}. 
Separate compilation supports modularity, incremental builds, and
third-party reuse.
The correctness requirement is simple: linking independently compiled
components should produce a program with the same behavior as linking
their sources before compilation. 
However, establishing this correspondence
is less routine than proving type soundness, even for simple types,
because the proof requires a formal account of linking.

\citet{Cardelli97} models separately compiled fragments as open
terms and links them by substitution within \emph{linksets}.
He proves that linking preserves typing; SMLSC states the same
property for Standard ML~\cite{SwaseyMCH2006}. Both link outside
the language, without proving that linked code behaves like the
combined source program.

Other designs bring linking into the language. Units supports
run-time composition~\cite{FlattFelleisen1998}, Alice ML expresses
linking through functions~\cite{RossbergMissingLink2006}, and MixML
provides recursive module linking~\cite{RossbergDreyerMixML2013}.
We know of no typed calculus with linking as an expression that
proves that linking separately compiled components preserves
source behavior.
However, verified compilers have established separate compilation correctness,
including SepCompCert~\cite{KangEtAlPOPL16} and
CertiCoq~\cite{CertiCoqICFP21}. For passes that commute with
linking, SepCompCert derives this result from whole-program
correctness. In both systems, however, linking remains outside
the source language, and neither theorem covers dynamic linking.

We bring the two strands together: linking within the language,
backed by a proof of separate compilation correctness that covers
both, dynamic and static linking.
Our approach is \emph{linking as evaluation}: a source link
elaborates into ordinary target terms so we can utilize the
semantics of the target to perform the linking. Thus, evaluation 
provides a module's imports and produces a linked library.

We develop this approach in $\sce$, a \emph{simply typed} calculus
with first-class modules, environments, and linking. First-class
environments \cite{TanOliveiraECOOP2023,lambdaeOOPSLA24} let programs
construct the environment in which a module runs, and use a linked
library or module as an environment for computation. $\sce$ elaborates
into $\lame$, a calculus of first-class environments \cite{lambdaeOOPSLA24}.
A separately compiled module with imports becomes a closed function
whose parameter supplies its import environment. Linking constructs
this environment from the required exports of another module, applies the function,
and combines their exports. Function application, field selection,
and environment composition thus perform the linking using the
target's existing evaluation rules.
Sandboxing restricts a module to its declared import: it cannot access
the scope where it is written and compiles to the same term wherever
it appears. Linking supplies only the selected import, without exposing
the scope at the link site. The client runs under its captured
environment extended with that import.

Programmers control linking: a link is an ordinary expression. The
same link can assemble compiled modules at top level, link against a module passed to a function at run time, or draw an
import from the current environment. The link's elaboration uses only its operands' core terms and
interfaces, so the core linker emits that same term from compiled fragments:
compiling and linking commute by construction. Commutation is thus
SepCompCert's easy case, holding by design and for dynamic links too.

We first prove that elaboration preserves types: well-typed source
links compile to well-typed core terms. Beyond the typing guarantees
studied by \citeauthor{Cardelli97}, we define $\sce$'s operational
semantics and prove that elaboration preserves source semantics. That is,
a source program and its elaboration evaluate to values related by
elaboration or diverge together. 
Since the core linker produces the same term as elaborating the
source link, semantics and behavior preservation establish separate
compilation correctness directly from $\sce$'s semantics. The result
covers static and dynamic linking, both at top level
and inside functions.
The static case (Corollary~\ref{thm:static-sc}) establishes,
for $\sce$, the semantic correctness property that
\citeauthor{Cardelli97} identifies but does not formalize.
We also prove \emph{interface abstraction}: a client's exports depend on
its provider only through the imported values.

Our contributions are:
\begingroup\listisep=0pt
\begin{itemize}\compactitems
\item \textbf{The $\sce$ calculus (\S\ref{sec:sce}).} A calculus
with \emph{first-class} modules, linking, and environments. We prove
its type safety.
\item \textbf{Type-directed elaboration to $\lame$ (\S\S\ref{sec:tde}--\ref{sec:separate-compilation}).}
We elaborate source programs into $\lame$, translating \emph{first-class}
modules and linking into ordinary target terms. We prove type and semantics
preservation.
\item \textbf{Separate compilation correctness (\S\ref{sec:separate-compilation}).}
We derive separate compilation correctness from semantics and behavior
preservation. The design of elaboration makes the proof simple.
We also prove interface abstraction and confinement for sandboxed
modules.
\item \textbf{Extensions (\S\ref{sec:extensions}).}
We extend the system with unions, iso-recursive types, fixpoints, and
multi-import linking, showing that the approach still applies with these
more expressive features.
\item \textbf{Mechanization and implementation (\S\ref{sec:implementation}).}
\footnote{The implementation and mechanization are available at \href{https://github.com/jam-khan/SCE-Lang}{https://github.com/jam-khan/SCE-Lang}}
We mechanize all results in Lean~4
\cite{10.1007/978-3-030-79876-5_37} and implement SCE-Lang,
a prototype compiler written in OCaml. 
It targets
WebAssembly~\cite{10.1145/3062341.3062363} with no linking instruction
and supports staged linking in both the core and WebAssembly.

\end{itemize}
\endgroup

 \section{Overview}\label{sec:overview}

This section shows, on small examples, how $\sce$ turns linking into an
expression evaluated under first-class environments. Compiling a link
then produces the same code as compiling its parts and linking them. We introduce first-class environments
(Section~\ref{sec:overview-environments}), and review \citeauthor{Cardelli97}'s account of linking and separate
compilation (Section~\ref{sec:overview-existing}). Next, we present the design
of $\sce$ (Section~\ref{sec:linking-inside}) and illustrate its use
through programming examples
(Section~\ref{sec:programming-with-linking}).

The examples use a simplified surface syntax of \textsf{SCE-Lang},
our implementation (Section~\ref{sec:implementation}).

\subsection{First-Class Environments}
\label{sec:overview-environments}

Environments are fundamental to language implementations: they hold the bindings against which code executes.
However, environments normally belong to the implementation, not the language, so programs cannot reach or manipulate them.
\emph{First-class
environments}~\cite{GelernterJL87,QueinnecDeRoure96}
lift this restriction: environments become ordinary values that
programs can build, inspect, extend, and pass to functions.
Because the body of a module is a sequence of bindings, a
first-class environment can also represent that body. $\sce$ builds on this
idea and elaborates into $\lame$~\cite{lambdaeOOPSLA24}, a call-by-value
calculus in which dependent merges~\cite{TanOliveiraECOOP2023} build
first-class environments.

\paragraph{Dependent merges.}
In $\lame$, a record $\lrec{l}{e}$ supplies a single labeled entry.
A merge $e_1 \mrg e_2$ evaluates its operands and joins their values
into one environment,
so multi-field records are merges of single-field ones. Merges are
\emph{dependent}: $e_1$ is evaluated first, and its value extends the
environment in which $e_2$ runs. In $\lame$ a typing context is the type of an environment, so the
typing rule extends the context $\Gamma$ with the left operand's type
and pairs the result types:
{\compactdisplayskips\[
\inferrule
  {\HasType{\Gamma}{e_1}{A} \\
   \HasType{\Gamma \mathbin{\&} A}{e_2}{B}}
  {\HasType{\Gamma}{e_1 \mrg e_2}{A \mathbin{\&} B}}
\]}
For example, the right component of
{\compactdisplayskips\[
\lrec{\code{x}}{2} \mrg \lrec{\code{y}}{\code{x} + 1}
\]}
refers to \code{x}. By the rule above, the merge has type
$\rcd{\code{x}}{\Int} \mathbin{\&} \rcd{\code{y}}{\Int}$. At run time,
the value $\lrec{\code{x}}{2}$ on the left is added to the environment before
the right component runs. So \code{x} resolves to $2$, and the merge
evaluates to $\lrec{\code{x}}{2} \mrg \lrec{\code{y}}{3}$.

Dependencies between declarations are pervasive in modules, where a definition
often refers to earlier definitions of the same module. A dependent
merge makes such a sequence of dependent declarations an ordinary expression.
The following declarations
\begin{lstlisting}
let x : Int = 2
let y : Int = x + 1
\end{lstlisting}
are modeled as the merge above, and its type gives the types of the
declarations.

\paragraph{Variables and closures.}
Every $\lame$ expression, such as the merge above, runs under a
current, or \emph{ambient}, environment.
To resolve a variable, $\lame$ looks it up in the ambient environment and
never substitutes into terms. Terms of $\lame$ are nameless: a variable is a
de Bruijn index~\cite{DEBRUIJN1972381} giving the position of an entry
in the environment. For readability, this section writes names instead of
indices. It also writes $\unit$ for both the empty environment and its type.
A function evaluates to a closure $\clos{v}{x{:}A}{e}$, which
pairs the function with the environment $v$ of its definition.

\paragraph{Programmer-controlled environments.}
Besides dependent merges, which build first-class environments, $\lame$ provides
constructs that let a program choose the environment a computation runs
in. The \emph{query} operator $\query$ returns the current environment. A
\emph{box} $e_1 \triangleright e_2$ runs $e_2$ under the local environment
produced by $e_1$, which replaces the ambient one. Assuming that the
ambient environment binds \code{x} to an integer, the expression
{\compactdisplayskips\[
(\query \mrg \lrec{\code{y}}{2}) \triangleright (\code{y} + \code{x})
\]}
is well typed. Here $\query$ returns the ambient environment, which
contains \code{x}. The merge extends it with the local binding $\code{y} = 2$. The box
then runs the body under the extended environment, so both \code{y}
and \code{x} are in scope. Dropping the query
gives
{\compactdisplayskips\[
\lrec{\code{y}}{2} \triangleright (\code{y} + \code{x})
\]}
which is ill typed, because the environment that the box supplies binds
only \code{y}. A box therefore runs a computation under an environment that
the program controls. Its body sees that environment's bindings and no others.
The simply typed $\lambda$-calculus has no such construct: a
$\lambda$ only extends the environment, so a function body sees every enclosing binding.

With the box construct, $\lame$ provides a form of sandboxed
modules~\cite{lambdaeOOPSLA24}. A module that takes other modules 
as arguments is boxed under the empty environment, so it acts only 
through its arguments. The treatment is informal:
$\lame$ has no module syntax, types or rules. Nor do
\citet{lambdaeOOPSLA24} study linking independently developed
modules or separate compilation correctness.

\subsection{Linking \emph{Outside} The Language}
\label{sec:overview-existing}

\begin{figure}[t]
\compactcaption
\noindent
\begin{minipage}[t]{0.47\linewidth}\vspace{0pt}
\srcfile{Add.scei}
\begin{srcnow}[remember as=sepproviderscei]{sceiground}{sepcomp-top}
\begin{lstlisting}[style=scei]
sig
  op : Int -> Int -> Int
  unit : Int
end
\end{lstlisting}
\end{srcnow}
\end{minipage}\hfill
\begin{minipage}[t]{0.47\linewidth}\vspace{0pt}
\srcfile{Double.scei}
\begin{srcnow}{sceiground}{sepcomp-top}
\begin{lstlisting}[style=scei]
sig double : Int -> Int end
\end{lstlisting}
\end{srcnow}
\end{minipage}

\smallskip
\noindent
\begin{minipage}[t]{0.47\linewidth}\vspace{0pt}
\srcfile{Add.sce}
\begin{srclater}{sceground}{sepcomp-bot}
\begin{lstlisting}[style=sce]
module Add = struct
  let op (x y : Int) : Int = x + y
  let unit : Int = 0
end
\end{lstlisting}
\end{srclater}
\end{minipage}\hfill
\begin{minipage}[t]{0.47\linewidth}\vspace{0pt}
\srcfile{Double.sce}
\begin{srcnow}[remember as=sepclient]{sceground}{sepcomp-bot}
\begin{lstlisting}[style=sce,escapechar=|]
|\srcmark{sepimport}|import Add
module Double = struct
  let double (x : Int) : Int = Add.op x x
end
\end{lstlisting}
\end{srcnow}
\end{minipage}\begin{tikzpicture}[remember picture, overlay]
  \draw[intfflow] (sepproviderscei.east)
    to[out=0, in=180] (sepclient.west |- sepimport);
\end{tikzpicture}

\smallskip
\hfill\srclegend
\caption{\code{Double.sce} compiled against interface \code{Add.scei}.}
\label{fig:sepcomp}
\compactfloatend
\end{figure}
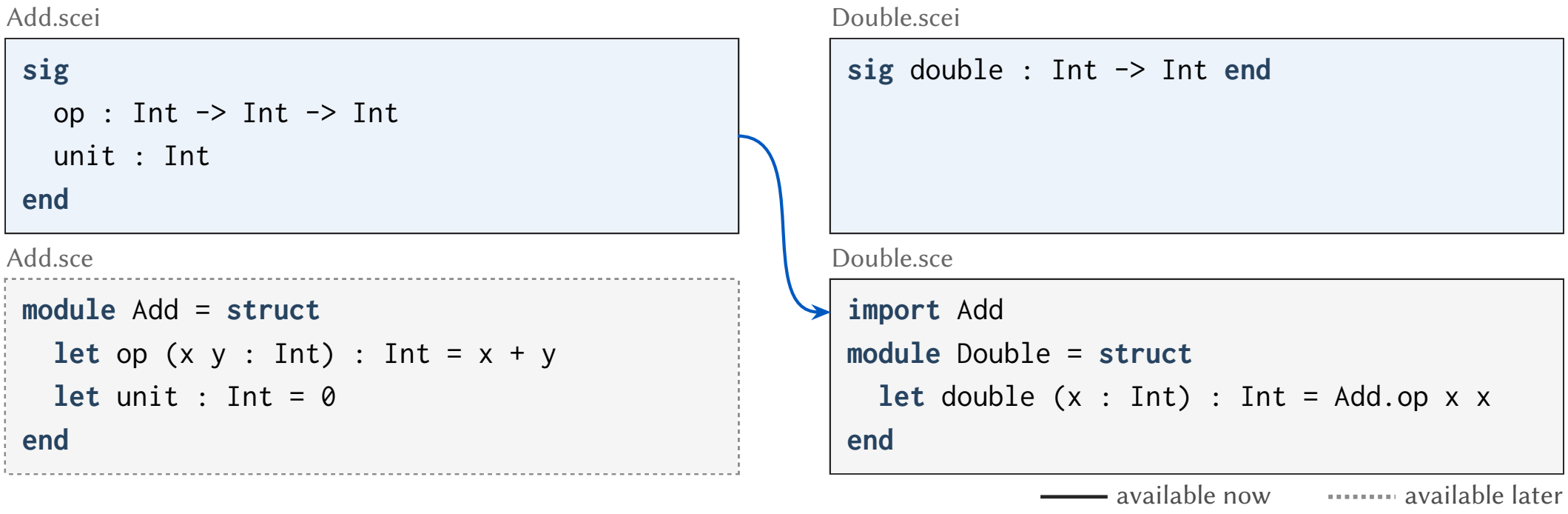

Separate compilation compiles a client module against its provider's
interface without the provider's implementation.
The provider can be compiled independently and linked with the compiled
client to produce an executable.
Source files have the extension \code{.sce}, and interface files, which
list a module's exports, have \code{.scei}. In Figure~\ref{fig:sepcomp},
\code{Double} imports \code{Add} and calls \code{Add.op} in
\code{double}. \code{Double.sce} takes the type of \code{Add.op} from
\code{Add.scei}, so it compiles before \code{Add.sce} is available. The figure shows only this
\emph{compilation} step, not how the modules are
\emph{linked} or what makes compiling and linking \emph{correct}.

\citet{Cardelli97} formalizes the linking step that the figure omits.
Separately compiled modules become \emph{linksets}: compilation
produces them, and combination and linking act on them. The underlying
calculus is $F_1$, a simply typed $\lambda$-calculus. A
judgment $E \vdash a : A$ assigns type $A$ to the term $a$ under the
typing environment $E$, a typing context rather than a run-time
environment. A linkset gathers judgments
under distinct names:
{\compactdisplayskips\[
E_0 \mid x_1 \lsto E_1 \vdash a_1 : A_1 \;\ldots\; x_n \lsto E_n \vdash a_n : A_n
\]}
The typing environment $E_0$ is the \emph{interface} of the linkset and lists
its imports. Each \emph{fragment} exports the term $a_i$ under the name
$x_i$. The typing environment $E_i$ of fragment $x_i$ gives the types of the fragments
that $a_i$ may refer to. Interfaces are thus typing environments outside the language: no term
of $F_1$ has a typing environment as its type.

\begin{figure}[t]
\compactcaption
\small
{\compactdisplayskips\[
\begin{array}{@{}r@{\;}c@{\;}l@{}}
\multicolumn{3}{@{}l@{}}{\lsstep{(a) Compilation}} \\[1pt]
\rlap{$L_{\code{Add}}$}\hphantom{L_{\code{Double}}} & = & \varnothing \mid
\begin{array}[t]{@{}l@{\;}c@{\;}l@{}}
\rlap{\code{op}}\hphantom{\code{double}} & \lsto & (\varnothing \vdash a_{+} : \Int \to \Int \to \Int) \\
\code{unit} & \lsto & (\varnothing,\ \code{op}{:}\Int \to \Int \to \Int \vdash 0 : \Int)
\end{array} \\[13pt]
L_{\code{Double}} & = & \varnothing,\ \code{op}{:}\Int \to \Int \to \Int \mid
\begin{array}[t]{@{}l@{\;}c@{\;}l@{}}
\code{double} & \lsto & (\varnothing \vdash \lambda x{:}\Int.\,\code{op}~x~x : \Int \to \Int)
\end{array} \\[7pt]
\multicolumn{3}{@{}l@{}}{\lsstep{(b) Combination}} \\[1pt]
L_{\code{Add}} + L_{\code{Double}} & = & \varnothing \mid
\begin{array}[t]{@{}l@{\;}c@{\;}l@{}}
\code{op} & \lsto & (\varnothing \vdash a_{+} : \Int \to \Int \to \Int) \\
\code{unit} & \lsto & (\varnothing,\ \code{op}{:}\Int \to \Int \to \Int \vdash 0 : \Int) \\
\code{double} & \lsto & (\varnothing,\ \lschg{\code{op}{:}\Int \to \Int \to \Int} \vdash \lambda x{:}\Int.\,\code{op}~x~x : \Int \to \Int)
\end{array} \\[26pt]
\multicolumn{3}{@{}l@{}}{\lsstep{(c) Linking}} \\[1pt]
\mathit{link}(L_{\code{Add}} + L_{\code{Double}}) & = & \varnothing \mid
\begin{array}[t]{@{}l@{\;}c@{\;}l@{}}
\code{op} & \lsto & (\varnothing \vdash a_{+} : \Int \to \Int \to \Int) \\
\code{unit} & \lsto & (\varnothing \vdash 0 : \Int) \\
\code{double} & \lsto & (\varnothing \vdash \lambda x{:}\Int.\,\lschg{a_{+}}~x~x : \Int \to \Int)
\end{array} \\[26pt]
\end{array}
\]}
\centering where $a_{+} = \lambda x{:}\Int.\lambda y{:}\Int.\,x + y$
\caption{Compiling, combining and linking the linksets of \code{Add} and \code{Double}, after \citet{Cardelli97}.}
\label{fig:linksets}
\compactfloatend
\end{figure}

Figure~\ref{fig:linksets} shows the three steps on the modules of
Figure~\ref{fig:sepcomp}, in adapted notation. Shading marks what each step changes. \citeauthor{Cardelli97}'s modules list imports one
variable at a time, not by module, so \code{Double} imports \code{op} directly. Compilation turns a
module into a linkset with one fragment per binding, each typed under the earlier bindings. The linkset's interface lists the
module's imports (a). Combining two linksets, written $L + L'$, puts their
fragments into one linkset. An import of one linkset that the other
exports moves from the interface to the typing environments of the
importing fragments (b). Linking, written $\mathit{link}(L)$, then
resolves every name in the typing environments by substitution: the
moved imports and the earlier bindings. Each such name leaves the
typing environment and is replaced by the term exported under that
name by a closed fragment, one typed under the empty environment (c).

\citeauthor{Cardelli97} proves that compilation, combination and linking preserve typing.
That is, each fragment typechecks, and each import has exactly the type
that its provider exports. \citeauthor{Cardelli97} names separate compilation correctness, but $F_1$ has no evaluation in which to state it. It holds when
linked code behaves like the merged sources, one program with both modules.
Linking acts on linksets, not on terms of $F_1$, so it lies outside the
language. A correctness proof thus needs an account of linking.

\citeauthor{Cardelli97}'s account raises three questions:
\begin{itemize}\compactitems
  \item Fragments are open terms, and linking rewrites them by
    substitution. Can a compiled module instead be a closed term that
    linking runs unchanged?
  \item Interfaces are typing environments, and linking is an operation
    on linksets. Can interfaces be ordinary types, and can linking be
    an expression evaluated during execution?
  \item Can separate compilation correctness follow from
    the compiler's correctness alone, with no account of linking?
\end{itemize}
Section~\ref{sec:linking-inside} answers them in order, using
first-class environments.

\subsection{Linking \emph{Inside} The Language}
\label{sec:linking-inside}

$\sce$ makes linking \emph{first-class}: a program can write a link at
top level, place it inside a function, or choose the link's provider at
run time. This design unifies static and dynamic linking in a
\emph{typed} setting. It also makes separate compilation correctness a
corollary of semantics and behavior preservation.

To illustrate, take \code{Add} and \code{Double} from
Figure~\ref{fig:sepcomp}, now declared within the program. We make
\code{Double} a \emph{sandboxed} functor requiring an \code{op} operation of type
$\Int \to \Int \to \Int$. Any provider exposing a unique \code{op}
of this type can be linked with \code{Double}.
First-class linking lets us combine these modules into a library
and use that library as the environment for a computation:
\par\nopagebreak\noindent\begin{minipage}{\linewidth}
\begin{lstlisting}
module Lib = link Add with Double
let main : Int = box Lib in op (double 5) unit    (* 10 *)
\end{lstlisting}
\end{minipage}\par
The \code{link} operator supplies \code{Double}'s import from
\code{Add} and combines their exports in \code{Lib}. Because structures
elaborate to \emph{first-class environments}, \code{Lib} serves
directly as the environment of \code{main}.

\paragraph{Modules as closed terms.}
\code{Lib} links two modules declared in the program. A separately
compiled file links in the same way, because a file with imports
elaborates to the same simplified form as a sandboxed functor, up to
parameter renaming. Figure~\ref{fig:module-elaboration}(a--b) shows this
form on a smaller module: a boxed lambda whose parameter binds the
imports. The box hides
the ambient environment and the lambda binds every import, so the
compiled file is closed. Linking applies it and never rewrites it,
unlike the substitution of Figure~\ref{fig:linksets}(c). This answers
the first question of Section~\ref{sec:overview-existing}.
In Figure~\ref{fig:module-elaboration}(c), a structure elaborates to a
first-class environment built by dependent merge: \code{y} can refer to
\code{x}.

\begin{figure}[t]
\compactcaption
\centering
\small
\begin{tabular}{@{}l@{\hspace{1.4em}}c@{\hspace{1.2em}}c@{}}
\multicolumn{3}{@{}l@{}}{\srcfile{(a) Importing file}} \\[2pt]
\begin{minipage}[c]{0.42\linewidth}
\begin{lstlisting}[style=srcfile]
import M (* sig x : Int; y : Int end *)
let z : Int = M.x
\end{lstlisting}
\end{minipage}
& $\rightsquigarrow$
& $\cterm{\begin{aligned}
&\unit\triangleright
  \bigl(\lambda \rcd{\code{M}}{\rcd{\code{x}}{\Int}\mathbin{\&}\rcd{\code{y}}{\Int}}.
  ~\lrec{\code{z}}{\underline{0}.\code{M}.\code{x}}\bigr)
\end{aligned}}$ \\[0.8\baselineskip]
\multicolumn{3}{@{}l@{}}{\srcfile{(b) Sandboxed functor}} \\[2pt]
\begin{minipage}[c]{0.42\linewidth}
\begin{lstlisting}[style=srcfile]
sandbox functor (I :
  { M : sig x : Int; y : Int end }) ->
  struct let z : Int = I.M.x end
\end{lstlisting}
\end{minipage}
& $\rightsquigarrow$
& $\cterm{\begin{aligned}
&\unit\triangleright
  \bigl(\lambda \rcd{\code{M}}{\rcd{\code{x}}{\Int}\mathbin{\&}\rcd{\code{y}}{\Int}}.
  ~\lrec{\code{z}}{\underline{0}.\code{M}.\code{x}}\bigr)
\end{aligned}}$ \\[0.8\baselineskip]
\multicolumn{3}{@{}l@{}}{\srcfile{(c) Structure}} \\[2pt]
\begin{minipage}[c]{0.42\linewidth}
\begin{lstlisting}[style=srcfile]
struct let x : Int = 1;
       let y : Int = x + 1 end
\end{lstlisting}
\end{minipage}
& $\rightsquigarrow$
& $\cterm{\lrec{\code{x}}{1}\mrg\lrec{\code{y}}{\underline{\code{x}} + 1}}$
\end{tabular}
\caption{Source modules (left) and simplified $\lame$ elaborations
(right, shaded gray). We omit administrative bindings.
In (a--b), $\underline{0}=\query.0$ is the import record, by
de Bruijn index. In (c), underlining marks a field lookup.}
\label{fig:module-elaboration}
\compactfloatend
\end{figure}

\paragraph{Interfaces as types.}
Module interfaces are ordinary types, so functions can accept and
return modules. This answers the first half of the second question. We
write $\Intf\;A$ for the type of a structure whose contents have type
$A$, and $A \marr B$ for a functor with import type $A$ and result type
$B$. The type translation $\elab{\cdot}$ maps the functor and structure
types in
Figure~\ref{fig:module-elaboration}(b--c) as follows:
{\compactdisplayskips\[
\begin{aligned}
\elab{
  \rcd{\code{M}}{\Intf\;(\rcd{\code{x}}{\Int}\mathbin{\&}\rcd{\code{y}}{\Int})}
  \marr \Intf\;\rcd{\code{z}}{\Int}}
&= \cterm{
  \rcd{\code{M}}{\rcd{\code{x}}{\Int}\mathbin{\&}\rcd{\code{y}}{\Int}}
  \to \rcd{\code{z}}{\Int}} \\[3pt]
\elab{\Intf\;(\rcd{\code{x}}{\Int}\mathbin{\&}\rcd{\code{y}}{\Int})}
&= \cterm{\rcd{\code{x}}{\Int}\mathbin{\&}\rcd{\code{y}}{\Int}}
\end{aligned}
\]}

\paragraph{Linking as evaluation.}
The second question also asks
whether linking can be an expression. In
$\sce$, a source link elaborates into ordinary $\lame$ terms, so the
core evaluator links with its existing rules for application, lookup, selection
and merge.

\par\smallskip\noindent
\textit{Link elaboration into $\lame$.}
\looseness=-1 Figure~\ref{fig:link-elaboration-example} elaborates the link of
\code{Add} (Figure~\ref{fig:sepcomp}) with \code{Double}, a sandboxed
functor (marked $\sandbox$) as in Figure~\ref{fig:module-elaboration}(b). We write
$\mlink{e_1}{e_2}$ for \code{link $e_1$ with $e_2$}.\footnote{The examples
assume integer arithmetic primitives, omitted from the formal grammar.}
The judgment $\Elab{\varepsilon}{e}{A}{e'}$ types a source term $e$ and
translates it to a core term $e'$. In \ruleref{Elab-Link}, the premise
$\lin{\code{op}}{\Int\to\Int\to\Int}{\Intf\;\code{P}}$ checks that
\code{Add} exports \code{op} exactly once, with the type \code{Double} imports.

\begin{figure}[t]
\compactcaption
\centering
\small
\setlength{\jot}{1pt}
\mprset{sep=1em, vskip=0.2ex}
\sbox{\formalbox}{$\displaystyle
\begin{gathered}
\lin{\code{op}}{\Int \to \Int \to \Int}{\Intf\;\code{P}} \\[-4pt]
\inferrule*[right=\rulerefx{Elab-Fun-sb}{Elab-Fun$_\sandbox$}]
  {\cdots}
  {\Elab{\varepsilon}{\code{Double}}{\code{D} \marr \Intf\;\code{T}}{n'}}
\qquad
\inferrule*[right=\ruleref{Elab-Str}]
  {\cdots}
  {\Elab{\varepsilon}{\code{Add}}{\Intf\;\code{P}}{m'}}
\end{gathered}
$}
{\compactdisplayskips\[
\begin{gathered}
\code{D} = \rcd{\code{op}}{\Int \to \Int \to \Int}
\quad
\code{P} = \code{D} \mathbin{\&} \rcd{\code{unit}}{\Int}
\quad
\code{T} = \rcd{\code{double}}{\Int \to \Int} \\[2pt]
\cterm{m'} = \cterm{
  \lrec{\code{op}}{\lambda x{:}\Int.\;\lambda y{:}\Int.\;x+y}
  \mrg \lrec{\code{unit}}{0}}\qquad
\cterm{n'} = \cterm{\unit \triangleright
  \bigl(\lambda \code{D}.\;
    \lrec{\code{double}}{\lambda x{:}\Int.\;\code{op}\;x\;x}\bigr)}
\\[3pt]
\inferrule*[right=\ruleref{Elab-Link}]
  {\usebox{\formalbox}}
  {\STyp{\varepsilon}{\mlink{\code{Add}}{\code{Double}}}
     {\Intf\;\code{P} \mathbin{\&} \Intf\;\code{T}}
   \rightsquigarrow\;
   \underbrace{\cterm{
     \bigl(\lambda p{:}\code{P}.\;\lambda f{:}(\code{D}\to\code{T}).\;
       p \mrg (f\;\lrec{\code{op}}{p.\code{op}})\bigr)\;m'\;n'
   }}_{e_\ell}}
\end{gathered}
\]}
\caption{Elaboration of \code{Add} linked with \code{Double}.}
\label{fig:link-elaboration-example}
\compactfloatend
\end{figure}

In $\cterm{e_\ell}$, the applications evaluate the provider and client once,
in that order, and bind their values to $p$ and $f$. The body selects
the \code{op} field from $p$ and supplies it to $f$ as
$\lrec{\code{op}}{p.\code{op}}$, then merges $p$ with the client's
result. Linking thus becomes ordinary application and
environment composition.
Because the current environment $\query$, which $\sce$ inherits from $\lame$, is a first-class value
(Section~\ref{sec:overview-environments}), it can also be the
provider. In a structure that declares \code{op}, with no other \code{op} in
scope, a later declaration can be $\mlink{\query}{\code{Double}}$, which
links \code{Double} against that \code{op}.

\paragraph{Confined linking.}
Linking against $\query$ makes the link site the provider, by explicit
choice. First-class environments invite a design that makes this
implicit: run the client under the environment where the link is
written, extended with its import. The client's meaning would then
depend on the link site. A name that the client uses but does not
import would resolve to whatever that site binds, and its compiled
code would be valid only there. In $\sce$, a link instead applies the
client, as $\cterm{e_\ell}$ shows. The client runs under the
environment its closure captured, extended with the selected import.
The link itself adds nothing from its own site. This holds for every
client. A
functor without the sandbox mark, an \emph{open functor}, still
captures the environment where it is written. Its body thus sees the
definition site. The box in $\cterm{n'}$ removes that environment as
well. \code{Double}'s body runs under the empty environment extended
with the record that $f$ receives, so it sees \code{op} from
\code{Add} and nothing else. A sandboxed client therefore compiles to
the same term wherever it is written. The compiled client runs at any
link site, against any provider that exports its import exactly once
at the required type, and its exports depend only on that import.
Context Independence (Lemma~\ref{lem:context-independence}) and
Sandbox and Link Confinement
(Theorems~\ref{thm:confinement} and~\ref{thm:link-confinement}) state
these properties.

\par\smallskip\noindent
\textit{Link typing and the core linker.}
Elaboration preserves typing (Section~\ref{sec:tde}), so a well-typed
source link elaborates to a well-typed core term. In the example, the source link has type
$\Intf\;\code{P} \mathbin{\&} \Intf\;\code{T}$:
$\code{P}$ describes \code{Add}'s exports (\code{op}, \code{unit}),
$\code{D}$ the import (\code{op}), and $\code{T}$ \code{Double}'s
export (\code{double}). Type translation gives the core types
$\code{P}$ for $\cterm{m'}$, $\code{D}\to\code{T}$ for $\cterm{n'}$,
and $\code{P}\mathbin{\&}\code{T}$ for $\cterm{e_\ell}$.

With these core types, the core linker needs no sources. Rule
\ruleref{CL-Link} checks that $\code{P}$ supplies the client's import
and emits $\cterm{e_\ell}$, the same term that
\ruleref{Elab-Link} produced:
{\compactdisplayskips\[
\inferrule*[right=\ruleref{CL-Link}]
  {\linc{\code{op}}{\Int\to\Int\to\Int}{\code{P}}}
  {\clink{\code{P}}{\cterm{m'}}
    {\code{D}\to\code{T}}{\cterm{n'}}{\cterm{e_\ell}}}
\]}
The rule does not typecheck $\cterm{m'}$ or $\cterm{n'}$. Given their
typings, which elaboration guarantees, Link typing
(Theorem~\ref{thm:core-link-typed}) derives
$\HasType{\varepsilon}{\cterm{e_\ell}}{\code{P}\mathbin{\&}\code{T}}$.

\par\smallskip\noindent
\textit{Linking semantics and preservation.}
A functor evaluates to a module closure $\mclos{v}{A}{e}$, which pairs
the functor with its captured environment $v$. Let $v_A$ be \code{Add}'s value, $c_+$ its addition
closure, and $b$ \code{Double}'s body. With $\rho = \unit \mrg \lrec{\code{op}}{c_+}$ and an empty ambient
environment, the following abridged reduction
supplies \code{op} to the client and retains both modules' exports:
{\compactdisplayskips\[
\begin{aligned}
\mlink{\code{Add}}{\code{Double}}
  &\longrightarrow^{*}
    \mlink{v_A}{\mclos{\unit}{\code{D}}{b}}
  \longrightarrow
    v_A \mrg (\rho \triangleright b) \\
  &\longrightarrow^{*}
    \underbrace{v_A \mrg
      \mstruct{\lrec{\code{double}}
        {\clos{\rho}{x{:}\Int}{\code{op}\;x\;x}}}}_{v_\ell}
\end{aligned}
\]}
The semantics of Section~\ref{sec:sstep} gives this reduction both for
separately compiled files linked at top level and for modules linked
inside a program.
Semantics preservation (Theorem~\ref{thm:semantics-preservation}) ensures
that $\cterm{e_\ell}$ evaluates to a value related to $v_\ell$ by elaboration.

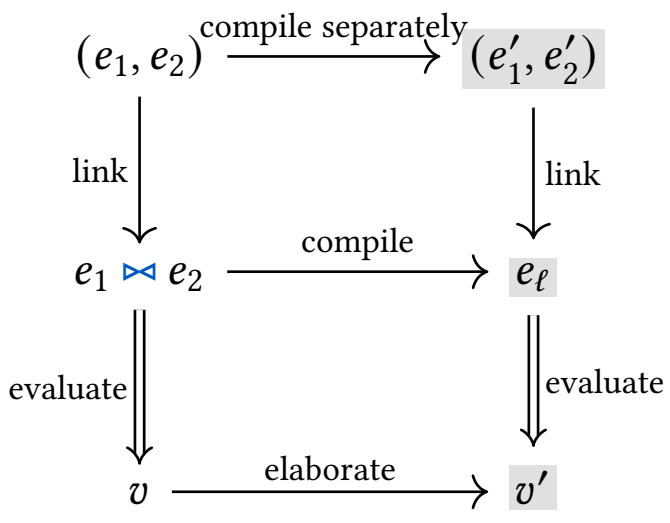
\begin{wrapfigure}{r}{0.35\linewidth}
\centering
\small
\begin{tikzcd}[column sep=4.5em, row sep=2.8em,
  /tikz/every label/.append style={font=\scriptsize}]
(e_1,e_2)
  \arrow[r, "\text{compile separately}"]
  \arrow[d, "\text{link}"']
& \cterm{(e_1',e_2')}
  \arrow[d, "\text{link}"] \\
\mlink{e_1}{e_2}
  \arrow[r, "\text{compile}"]
  \arrow[d, Rightarrow, "\text{evaluate}"']
& \cterm{e_\ell}
  \arrow[d, Rightarrow, "\text{evaluate}"] \\
v
  \arrow[r, "\text{elaborate}"]
& \cterm{v'}
\end{tikzcd}
\caption{Correctness diagram.}
\Description{Two commuting squares, with core terms shaded gray.
The upper square compares compiling a source link with independently
compiling and then linking its operands. Both routes produce the same
core term. The lower square relates terminating source evaluation to
core evaluation through elaboration of the resulting values.}
\label{fig:overview-correctness}
\end{wrapfigure}
\paragraph{Separate compilation correctness.}
We can now answer the last question:
separate compilation correctness is a corollary of semantics and behavior preservation.
Table~\ref{tab:cardelli} summarizes all three answers.

The upper square of Figure~\ref{fig:overview-correctness} compares the
two paths. Rule \ruleref{Elab-Link} elaborates the source link. That link is
the merged program, and the source semantics gives its meaning. Rule \ruleref{CL-Link} links
the independently compiled operands using only their interfaces. Both
paths produce the same core term $\cterm{e_\ell}$, so elaboration
commutes with linking by construction. The
lower square is semantics preservation, for terminating runs. With behavior preservation
(Theorem~\ref{thm:behavior-preservation}), it yields separate
compilation correctness for every well-typed link, at top level
or in a program (Corollaries~\ref{thm:separate-compilation}
and~\ref{thm:static-sc}).

\begin{table}[t]
\compactcaption
\centering
\caption{Separate compilation with linksets and $\sce$.}
\label{tab:cardelli}
\begin{tabular}{@{}l@{\qquad}l@{\qquad}l@{}}
\toprule
                    & Linksets \cite{Cardelli97} & $\sce$               \\
\midrule
Compiled modules     & Open fragments        & Closed terms                       \\
Interfaces           & Typing environments   & Ordinary types                     \\
Linking              & Operation on linksets & Ordinary expression                \\
Import resolution    & Term substitution     & Argument passing and lookup        \\
Separate compilation & Type preservation     & Type, semantics and behavior preservation    \\
\bottomrule
\end{tabular}
\compactfloatend
\end{table}

\FloatBarrier
\subsection{Programming with \emph{First-Class} Linking and Environments}
\label{sec:programming-with-linking}

Linkset linking (Section~\ref{sec:overview-existing}) runs before
execution, so it cannot take a provider that a program builds at run
time. In $\sce$, a link is an expression. So a program can load a
compiled client at run time and link it against a provider built per
call. Figure~\ref{fig:plugin-host} builds a page renderer whose formatting
plugin is such a client. The figure omits the imports of \code{Sys} and
\code{Loader} and the union declarations. The host supplies logging and falls back to
plain text if loading fails.

\begin{figure}[t]
\compactcaption
\centering
\begin{minipage}{0.94\linewidth}
\srcfile{(a) Compile the plugin separately: fancy.sce to fancy.sceo}
\begin{lstlisting}[style=srcfile]
import Cap : { log : String -> Unit }
let render (text : String) : String =
  let noted : Unit = Cap.log "rendering" in
  "** " ^ text ^ " **"
\end{lstlisting}

\medskip
\srcfile{(b) Build a logger, then package it with a message}
\begin{lstlisting}[style=srcfile]
let caps (name : String) =
  { log = fun (s : String) -> Sys.print ("[" ^ name ^ "] " ^ s) }
let host (cap : { log : String -> Unit }) (text : String) = struct
  let message : String = text
  let Cap = cap
end
\end{lstlisting}

\medskip
\srcfile{(c) Load the client, link it, and run under the resulting environment}
\begin{lstlisting}[style=srcfile]
let page (cap : { log : String -> Unit })
         (path text : String) : String =
  let saved = host cap text in
  match Loader.load path with
  | Plugin p ->
      let lib = link saved with p in
      box lib in render message
  | Failed e -> box saved in message
  end
\end{lstlisting}

\medskip
\begin{lstlisting}[style=srcfile]
let aliceCap = caps "alice"
page aliceCap "fancy.sceo" "hello"
  (* prints "[alice] rendering"; returns "** hello **" *)
page aliceCap "fancy.sceo" "bye"
  (* prints "[alice] rendering"; returns "** bye **" *)
page aliceCap "missing.sceo" "hello"  (* returns "hello" *)
\end{lstlisting}
\end{minipage}
\caption{A plugin host in three steps. Comments show the output.}
\Description{The plugin imports a logging capability and exports a renderer.
The host builds a provider with a message and a caller-specific logger.
It loads a plugin and runs under the linked environment on success,
or under the saved provider on failure. Two successful calls reuse the
same logger with different messages. A missing plugin falls back to plain text.}
\label{fig:plugin-host}
\compactfloatend
\end{figure}

\paragraph{Compile the client once.}
First, the plugin in panel~(a) imports \code{Cap} for logging and exports
\code{render}. Compilation produces a sandboxed functor. Its body can call
\code{Cap.log} but cannot name the host's \code{Sys}, although
\code{Sys} is in scope where the host links it
(Section~\ref{sec:linking-inside}). Sandboxed functors
elaborate independently of their ambient context
(Lemma~\ref{lem:context-independence}).

\paragraph{Construct the provider.}
Next, \code{caps} in panel~(b) builds a logger that captures the caller's
\code{name}, and \code{host} packages it with a message. Both successful
calls reuse \code{aliceCap}, although their providers carry different
messages. Linking passes only \code{Cap} to the plugin, which exports
the same \code{render} closure in both cases. This illustrates interface
abstraction (Theorem~\ref{thm:interface-abstraction}).
The linked result still contains \code{message}. A no-op logger would
silence the plugin without recompiling it.

\paragraph{Link, then choose the environment.}
Finally, \code{page} in panel~(c) loads the plugin. On success, linking
\code{saved} with \code{p} produces \code{lib}. The box evaluates
\code{render message} under \code{lib}, using the plugin's function and
the provider's message. The calls return different strings because the
host supplies different messages to that function. On failure, the box
reads the message from the saved provider.

The declared type of \code{Loader.load} gives \code{p} the type of
panel~(a), so the compiler checks the link before loading. At run time,
\code{Loader.load} returns \code{Plugin p} only for a compiled file
of that type.

The example illustrates the linking construction covered by separate
compilation correctness (Corollary~\ref{thm:separate-compilation}).
Strings, I/O, and artifact loading lie outside the formal calculus.

\FloatBarrier
 \section{The Source Calculus: \texorpdfstring{$\sce$}{lambda-SCE}}\label{sec:sce}

\begingroup\widowpenalty=0
$\sce$ extends $\lame$~\cite{lambdaeOOPSLA24} with
first-class modules and a linking expression. It keeps first-class environments and dependent merges~\cite{TanOliveiraECOOP2023}. The merged program is the source link, which $\sce$'s semantics evaluates: Section~\ref{sec:separate-compilation} states separate compilation against that evaluation. Rule \ruleref{Elab-Link} translates a link to the term that the core linker emits for compiled fragments. By Corollary~\ref{cor:type-safety}, a well-typed program cannot get stuck, even for links in functions or with providers chosen at run time.\par\endgroup

\begin{figure}[t]
\compactcaption
\centering
\syntaxrow{
\begin{array}{@{}l@{\qquad}rcl@{}}
\syntaxcat{Types} & A, B, \Gamma & ::= & \Int \mid \varepsilon \mid A \to B \mid A \mathbin{\&} B \mid \rcd{l}{A} \mid \Intf\; A \mid A \marr B \\
\syntaxcat{Functor modes} & s & ::= & \sandbox \mid \open \\
\syntaxcat{Expressions} & e & ::= & \query \mid i \mid \unit \mid e.n \mid e.l \mid \lambda A.\; e \mid \clos{v}{A}{e} \mid e_1\;e_2 \mid \lrec{l}{e} \\
 & & \mid & e_1\triangleright e_2 \mid e_1\mrg e_2 \mid e_1\nmrg e_2 \mid \letb{e_1}{e_2} \mid \openm{e_1}{e_2} \mid \modl{m} \\
\syntaxcat{Modules} & m & ::= & \mstruct{e} \mid \mfunctor{s}{A}{e} \mid \mclos{v}{A}{e} \mid \mapp{e_1}{e_2} \mid \mlink{e_1}{e_2} \\
\syntaxcat{Values} & v & ::= & i \mid \unit \mid \clos{v}{A}{e} \mid \mclos{v}{A}{e} \mid \mstruct{v} \mid v_1\mrg v_2 \mid \lrec{l}{v} \\
\syntaxcat{Sugar} & \underline{l} & \triangleq & \query.l \qquad\qquad \underline{n}\triangleq\query.n
\end{array}
}
\caption{Syntax of $\sce$.}
\label{fig:sce-syntax}
\compactfloatend
\end{figure}

\subsection{Syntax}\label{sec:scesyntax}
Figure~\ref{fig:sce-syntax} shows the syntax of $\sce$. We use
\textcolor{modcolor}{blue} for modules and module types, and
\textcolor{extcolor}{orange} for the other extensions to $\lame$.
The black constructs are inherited from $\lame$. These colors also
distinguish the constructs in the typing and reduction rules.

\subsubsection*{Types as contexts.} Metavariables $A$, $B$, and
$\Gamma$ range over types. Any type can be a typing context.
Types include integers ($\Int$), a unit type ($\varepsilon$) for
the empty environment, environment types ($A \mathbin{\&} B$),
arrow types ($A \to B$), and single-field record types
($\rcd{l}{A}$). An environment type pairs
two environment components in order. There is no subtyping. Type equality is structural, so $\Int \mathbin{\&} \varepsilon$ differs from $\varepsilon \mathbin{\&} \Int$, as $(\Int \mathbin{\&} \Int) \mathbin{\&} \varepsilon$ does from $\Int \mathbin{\&} (\Int \mathbin{\&} \varepsilon)$. We associate chains of environment types and merges to the
left. A multi-field record type abbreviates a chain of single-field ones.

\subsubsection*{Module types.} Any module's type is its interface: an interface type for a structure, and a functor type for a functor. The interface type $\Intf\; A$, written with the keyword \code{sig}, classifies structures whose body has type $A$. The functor type $A \marr B$ classifies functors
importing $A$ and exporting $B$.

\subsubsection*{Expressions.} Expressions include integers ($i$), a unit expression ($\unit$), applications ($e_1\; e_2$), boxes ($e_1 \triangleright e_2$), dependent merges ($e_1 \mrg e_2$), and lambda abstractions ($\lambda A.\; e$). A lambda binder is nameless. Its annotation $A$ supplies the parameter type when typing the body. The query $\query$ returns the current environment. Projection ($e.n$) returns the $n$-th entry of the value of $e$, so a variable with de Bruijn index~$n$~\cite{DEBRUIJN1972381} is $\query.n$, abbreviated $\underline{n}$. Records ($\lrec{l}{e}$) create labeled entries, which selection ($e.l$) retrieves. Multi-field records are merges of single-field ones. The sugar $\underline{l}$ abbreviates $\query.l$. A closure ($\clos{v}{A}{e}$) saves the environment in which a function was created.

The orange constructs elaborate to core expressions
(Section~\ref{sec:elaboration}). A non-dependent merge ($e_1 \nmrg
e_2$) evaluates both components under the same environment, without
making the value of $e_1$ available to $e_2$.
The let binding ($\letb{e_1}{e_2}$) adds the value of $e_1$ to the
environment of $e_2$. It requires no annotation because typing
synthesizes the type of $e_1$ (Section~\ref{sec:scetyping}). Finally,
$\openm{e_1}{e_2}$ opens a named component: $e_1$ must have a
single-field record type $\rcd{l}{A}$, and its field value extends
the environment of $e_2$. If the field holds a structure, $e_2$ can then select the structure's exports by label.

\subsubsection*{Modules.} The blue constructs form the syntactic category $m$ of modules, which is part of the expression grammar ($e ::= \cdots \mid m$). Their subterms are arbitrary
expressions. In particular, the operands of module application and
linking need not be module forms. A functor may be selected from the
environment or computed by another expression.

A structure ($\mstruct{e}$) evaluates its body under the ambient environment and packages the result as the value $\mstruct{v}$. A functor
($\mfunctor{s}{A}{e}$) abstracts a body over an import of type $A$.
Like a lambda abstraction, a functor evaluates to a closure, here a module
closure ($\mclos{v}{A}{e}$). The functor's mode $s$ determines the captured
environment: empty in sandbox mode ($\sandbox$) and ambient in open
mode ($\open$). Module application
($\mapp{e_1}{e_2}$) supplies the functor $e_1$ with its import
$e_2$. Modules can be bound, merged, passed to functions, and
returned.

Linking ($\mlink{e_1}{e_2}$) connects a provider $e_1$ to a client functor $e_2$. The client's import is a single-field record
$\rcd{l}{A}$, as for $\kwe{open}$, and its label $l$ names an export of
the provider.
Linking selects that export, supplies it to the client, and merges the provider with the client's result. The provider can be any expression whose type contains the import, such as a function's argument or $\query$ (\ruleref{Elab-Env}).

\subsection{Typing}\label{sec:scetyping}

Figure~\ref{fig:elaboration-core} presents the typing and elaboration
rules for expressions. The judgment $\Elab{\Gamma}{e}{A}{e'}$ states that
source expression $e$ has type $A$ under $\Gamma$ and
elaborates to the $\lame$ term $e'$. Elaborated terms are shaded
gray, and $\elab{A}$ translates a source type to a
$\lame$ type. This subsection explains typing. Section~\ref{sec:sstep} explains the term that rule \ruleref{Elab-Link} generates, and Section~\ref{sec:tde} the type translation and other generated terms.

\begin{figure*}[t]
\compactcaption
\centering
\begin{rulefigure}
\ruleheading{\lin{l}{A}{B}}{Containment}
\rulerow{
  \inferrule*[lab=\rulelabel{Ctm-Rcd}]
    { }
    {\lin{l}{A}{\rcd{l}{A}}}
  \quad
  \inferrule*[lab=\rulelabel{Ctm-AndL}]
    {\lin{l}{A}{B} \\ l \notin \labels{C}}
    {\lin{l}{A}{B \mathbin{\&} C}}
  \quad
  \inferrule*[lab=\rulelabel{Ctm-AndR}]
    {\lin{l}{A}{C} \\ l \notin \labels{B}}
    {\lin{l}{A}{B \mathbin{\&} C}}
  \quad
  \inferrule*[lab=\rulelabel{Ctm-Sig}]
    {\lin{l}{A}{B}}
    {\lin{l}{A}{\Intf\; B}}}
\ruleheading{\Elab{\Gamma}{e}{A}{e'}}{Typing and elaboration}
\rulerow{
  \inferrule*[lab=\rulelabel{Elab-Int}]
    { }
    {\Elab{\Gamma}{i}{\Int}{i}}
  \quad
  \inferrule*[lab=\rulelabel{Elab-Top}]
    { }
    {\Elab{\Gamma}{\unit}{\varepsilon}{\unit}}
  \quad
  \inferrule*[lab=\rulelabel{Elab-Env}]
    { }
    {\Elab{\Gamma}{\query}{\Gamma}{\query}}
  \quad
  \inferrule*[lab=\rulelabel{Elab-Idx}]
    {\Elab{\Gamma}{e}{A}{e'} \\ \lookup(A, n) = B}
    {\Elab{\Gamma}{e.n}{B}{e'.n}}}
\rulerow{
  \inferrule*[lab=\rulelabel{Elab-Lam}]
    {\Elab{\Gamma \mathbin{\&} A}{e}{B}{e'}}
    {\Elab{\Gamma}{\lambda A.\; e}{A \to B}{\lambda \elab{A}.\; e'}}
  \quad
  \inferrule*[lab=\rulelabel{Elab-Rcd}]
    {\Elab{\Gamma}{e}{A}{e'}}
    {\Elab{\Gamma}{\lrec{l}{e}}{\rcd{l}{A}}{\lrec{l}{e'}}}
  \quad
  \inferrule*[lab=\rulelabel{Elab-Sel}]
    {\Elab{\Gamma}{e}{B}{e'} \\
     \lin{l}{A}{B}}
    {\Elab{\Gamma}{e.l}{A}{e'.l}}}
\rulerow{
  \inferrule*[width=0.31\linewidth,lab=\rulelabel{Elab-App}]
    {\Elab{\Gamma}{e_1}{A \to B}{e_1'} \\
     \Elab{\Gamma}{e_2}{A}{e_2'}}
    {\Elab{\Gamma}{e_1\; e_2}{B}{e_1'\; e_2'}}
  \qquad
  \inferrule*[width=0.31\linewidth,lab=\rulelabel{Elab-Box}]
    {\Elab{\Gamma}{e_1}{\Gamma_1}{e_1'} \\
     \Elab{\Gamma_1}{e_2}{A}{e_2'}}
    {\Elab{\Gamma}{e_1 \triangleright e_2}{A}{e_1' \triangleright e_2'}}
  \qquad
  \inferrule*[width=0.31\linewidth,lab=\rulelabel{Elab-Mrg}]
    {\Elab{\Gamma}{e_1}{A}{e_1'} \\
     \Elab{\Gamma \mathbin{\&} A}{e_2}{B}{e_2'}}
    {\Elab{\Gamma}{e_1 \mrg e_2}{A \mathbin{\&} B}{e_1' \mrg e_2'}}}
\rulerow{
  \inferrule*[lab=\rulelabel{Elab-Clos}]
    {\Elab{\varepsilon}{v}{\Gamma_1}{v'} \\
     \Elab{\Gamma_1 \mathbin{\&} A}{e}{B}{e'}}
    {\Elab{\Gamma}{\clos{v}{A}{e}}{A \to B}{\clos{v'}{\elab{A}}{e'}}}
  \quad
  \inferrule*[lab=\rulelabel{Elab-NMrg}]
    {\Elab{\Gamma}{e_1}{A}{e_1'} \\
     \Elab{\Gamma}{e_2}{B}{e_2'}}
    {\STyp{\Gamma}{e_1 \nmrg e_2}{A \mathbin{\&} B} \\
     \rightsquigarrow\; \cterm{\bigl(\query \mrg
        ((\underline{0} \triangleright e_1') \mrg
         (\underline{1} \triangleright e_2'))\bigr).0}}}
\rulerow{
  \inferrule*[lab=\rulelabel{Elab-Let}]
    {\Elab{\Gamma}{e_1}{A}{e_1'} \\
     \Elab{\Gamma \mathbin{\&} A}{e_2}{B}{e_2'}}
    {\Elab{\Gamma}{\letb{e_1}{e_2}}{B}{(e_1' \mrg e_2').0}}
  \qquad
  \inferrule*[lab=\rulelabel{Elab-Open}]
    {\Elab{\Gamma}{e_1}{\rcd{l}{A}}{e_1'} \\
     \Elab{\Gamma \mathbin{\&} A}{e_2}{B}{e_2'}}
    {\Elab{\Gamma}{\openm{e_1}{e_2}}{B}{(\lambda \elab{A}.\; e_2')\;
(e_1'.l)}}}
\end{rulefigure}
\caption{Containment, typing, and elaboration of expressions.}
\label{fig:sce-lookup}
\label{fig:elaboration-core}
\compactfloatend
\end{figure*}

\begin{figure*}[t]
\compactcaption
\centering
\begin{rulefigure}
\ruleheading{\Elab{\Gamma}{e}{A}{e'}}{Typing and elaboration}
\rulerow{
  \inferrule*[lab=\rulelabel{Elab-Str}]
    {\Elab{\Gamma}{e}{B}{e'}}
    {\Elab{\Gamma}{\mstruct{e}}{\Intf\; B}{e'}}
  \qquad
  \inferrule*[lab=\rulelabel{Elab-MApp}]
    {\Elab{\Gamma}{e_1}{A \marr B}{e_1'} \\
     \Elab{\Gamma}{e_2}{A}{e_2'}}
    {\Elab{\Gamma}{\mapp{e_1}{e_2}}{B}{e_1'\; e_2'}}}
\rulerow{
  \inferrule*[lab=\rulelabelx{Elab-Fun-sb}{Elab-Fun$_\sandbox$}]
    {\Elab{\varepsilon \mathbin{\&} A}{e}{B}{e'}}
    {\Elab{\Gamma}{\mfunctor{\sandbox}{A}{e}}{A \marr B}{\unit
\triangleright (\lambda \elab{A}.\; e')}}
  \qquad
  \inferrule*[lab=\rulelabelx{Elab-Fun-op}{Elab-Fun$_\open$}]
    {\Elab{\Gamma \mathbin{\&} A}{e}{B}{e'}}
    {\Elab{\Gamma}{\mfunctor{\open}{A}{e}}{A \marr B}{\lambda
\elab{A}.\; e'}}}
\rulerow{
  \inferrule*[lab=\rulelabel{Elab-MClos}]
    {\Elab{\varepsilon}{v}{\Gamma_1}{v'} \\
     \Elab{\Gamma_1 \mathbin{\&} A}{e}{B}{e'}}
    {\Elab{\Gamma}{\mclos{v}{A}{e}}{A \marr
B}{\clos{v'}{\elab{A}}{e'}}}}
\rulerow{
  \inferrule*[lab=\rulelabel{Elab-Link}]
    {\Elab{\Gamma}{e_1}{\Gamma_1}{e_1'} \\
     \Elab{\Gamma}{e_2}{\rcd{l}{A} \marr B}{e_2'} \\
     \lin{l}{A}{\Gamma_1}}
    {\STyp{\Gamma}{\mlink{e_1}{e_2}}{\Gamma_1 \mathbin{\&} B} \\
     \rightsquigarrow\; \cterm{
       (\lambda \elab{\Gamma_1}.\;
        \lambda (\rcd{l}{\elab{A}} \to \elab{B}).\;
        \underline{1} \mrg (\underline{1}\;\lrec{l}{\underline{0}.l}))\;
       e_1'\;e_2'}}}
\end{rulefigure}
\caption{Typing and elaboration of modules.}
\label{fig:elaboration-module}
\compactfloatend
\end{figure*}

\subsubsection*{Environments and lookup.}
The query $\query$ inspects the context, and has the current context
$\Gamma$ as its type (\ruleref{Elab-Env}). A projection $e.n$ has type
$\lookup(A,n)$ when $e$ has type $A$ (\ruleref{Elab-Idx}). The partial function $\lookup$ counts from the right end, so index $0$ is the rightmost entry:
{\compactdisplayskips\[
\begin{array}{@{}rcl@{\qquad\quad}rcl@{}}
\lookup(A \mathbin{\&} B, 0) &=& B
  & \lookup(A \mathbin{\&} B, n{+}1) &=& \lookup(A, n)
\end{array}
\]}
A box $e_1 \triangleright e_2$ replaces the ambient context $\Gamma$
with the type of $e_1$ when checking $e_2$ (\ruleref{Elab-Box}).
Lambdas and dependent merges retain $\Gamma$ and extend it on the
right to $\Gamma \mathbin{\&} A$, where $A$ is the parameter type
or the left operand's type, respectively.
By rule \ruleref{Elab-Clos}, a closure $\clos{v}{A}{e}$ uses its captured environment: its body is checked under $\Gamma_1 \mathbin{\&} A$, where $v$ has type $\Gamma_1$ under $\varepsilon$.

\subsubsection*{Labels and containment.}
Projection finds an environment entry by position. Selection and linking instead find a field by label. Both use the containment judgment $\lin{l}{A}{B}$ of Figure~\ref{fig:sce-lookup}: type $B$ exposes exactly one field named $l$, and that field has type $A$. The function
$\labels{\cdot}$ records the labels exposed by a type:
{\compactdisplayskips\[
\begin{array}{@{}rcl@{\qquad\quad}rcl@{}}
\labels{\rcd{l}{A}} &=& \{l\}
  & \labels{A \mathbin{\&} B} &=& \labels{A} \cup \labels{B} \\[2pt]
\labels{\Intf\; A} &=& \labels{A}
  & \labels{A} &=& \varnothing \quad\text{otherwise}
\end{array}
\]}
For an environment type, containment requires $l$ in exactly one
component.
A merge may contain duplicate labels, but selecting a duplicated label or linking a client that imports it is ill typed.

\subsubsection*{Let and open.}
Like selection, $\kwe{open}$ reads a field by its label, but it needs no containment. Its operand $e_1$ must have exactly the type $\rcd{l}{A}$. Rule \ruleref{Elab-Open} adds only $A$ to the context. Rule \ruleref{Elab-Let} instead adds the whole synthesized type of $e_1$, so for the same $e_1$ it adds $\rcd{l}{A}$. Take $e_1 = \lrec{\code{M}}{\mstruct{\lrec{\code{x}}{1}}}$. After $\openm{e_1}{e_2}$, $e_2$ selects the structure's export as $\underline{\code{x}}$. After $\letb{e_1}{e_2}$, $e_2$ writes $\underline{\code{M}}.\code{x}$. Either way, the first label $e_2$ selects, \code{x} or \code{M}, must be unique in the context.

\subsubsection*{Modules and linking.}
Figure~\ref{fig:elaboration-module} gives the module rules.
Like a $\kwe{let}$ or $\kwe{open}$ body, a functor body is checked under an extended context. Its import type $A$ extends the empty context or the ambient one, depending on the functor's mode. A sandboxed body is checked under $\varepsilon \mathbin{\&} A$, so it accesses only its import. An open body is checked under $\Gamma \mathbin{\&} A$ and may also use ambient bindings. Both modes give the functor the type $A \marr B$, where $B$ is the type of its body. Open mode serves functors over ambient modules. Sandbox mode serves separately compiled clients.

Rule \ruleref{Elab-Link} constructs the import from the one provider field of the required name and type. Although $\sce$ has no subtyping, linking thus accepts a provider with additional exports. Module application returns only the client's result,
whereas linking returns the provider merged with that result. For example, take \code{Add} from Figure~\ref{fig:sepcomp}, which exports $\code{op}:\Int\to\Int\to\Int$ and $\code{unit}:\Int$, and a functor $\code{Double}'$ that imports only $\rcd{\code{op}}{\Int\to\Int\to\Int}$ and exports $\code{double}:\Int\to\Int$.
Direct application $\mapp{\code{Double}'}{\code{Add}}$ is ill typed, because \code{Add}'s interface differs from the required import record type. The adapted application $\mapp{\code{Double}'}{\lrec{\code{op}}{\code{Add}.\code{op}}}$ returns only the client's structure, whereas $\mlink{\code{Add}}{\code{Double}'}$ needs no adapter. The link also retains \code{op} and \code{unit}.

\subsection{Small-Step Semantics}\label{sec:sstep}
Figure~\ref{fig:sstep} defines the call-by-value small-step semantics
of $\sce$. The judgment $\SStep{v}{e}{e'}$ states that $e$ takes
one step to $e'$ under the ambient environment $v$, which is a value.
We write $\SMStep{v}{e}{e'}$ for its reflexive and transitive closure.
\begin{figure*}[t]
\compactcaption
\centering
\begin{rulefigure}
{\noindent\formalrow{
\begin{array}{@{}l@{\qquad}rcl@{}}
\syntaxcat{Frames} & F & ::= & \hole.n \mid \hole.l \mid \lrec{l}{\hole} \mid \hole\;e \mid v\;\hole \mid \hole\triangleright e \mid \hole\mrg e \\
 & & \mid & \hole\nmrg e \mid v\nmrg\hole \mid \letb{\hole}{e} \mid \openm{\hole}{e} \\
 & & \mid & \mstruct{\hole} \mid \mapp{\hole}{e} \mid \mapp{v}{\hole} \mid \mlink{\hole}{e} \mid \mlink{v}{\hole} \\
\syntaxcat{Environment} & \ambm{\sandbox}{v} & = & \unit \qquad\qquad \ambm{\open}{v}\;=\;v
\end{array}
}\par}
\ruleheading{\selj{v}{l}{v'}}{Selection}
\rulerow{
  \inferrule*[lab=\rulelabel{Sel-Rcd}]
    { }
    {\selj{\lrec{l}{v}}{l}{v}}
  \quad
  \inferrule*[lab=\rulelabel{Sel-MrgL}]
    {\selj{v_1}{l}{v}}
    {\selj{(v_1 \mrg v_2)}{l}{v}}
  \quad
  \inferrule*[lab=\rulelabel{Sel-MrgR}]
    {\selj{v_2}{l}{v}}
    {\selj{(v_1 \mrg v_2)}{l}{v}}
  \quad
  \inferrule*[lab=\rulelabel{Sel-Str}]
    {\selj{v_1}{l}{v}}
    {\selj{\mstruct{v_1}}{l}{v}}}
\ruleheading{\SStep{v}{e}{e'}}{Small-step reduction}
\rulerow{
  \inferrule*[lab=\rulelabel{S-Frame}]
    {\SStep{v}{e}{e'}}
    {\SStep{v}{F[e]}{F[e']}}
  \enspace
  \inferrule*[lab=\rulelabel{S-BoxR}]
    {\SStep{v_1}{e_2}{e_2'}}
    {\SStep{v}{v_1 \triangleright e_2}{v_1 \triangleright e_2'}}
  \enspace
  \inferrule*[lab=\rulelabel{S-MrgR}]
    {\SStep{v \mrg v_1}{e_2}{e_2'}}
    {\SStep{v}{v_1 \mrg e_2}{v_1 \mrg e_2'}}}
\rulerow{
  \inferrule*[lab=\rulelabel{S-Env}]
    { }
    {\SStep{v}{\query}{v}}
  \quad
  \inferrule*[lab=\rulelabel{S-BoxV}]
    { }
    {\SStep{v}{v_1 \triangleright v_2}{v_2}}
  \quad
  \inferrule*[lab=\rulelabel{S-Lam}]
    { }
    {\SStep{v}{\lambda A.\; e}{\clos{v}{A}{e}}}
  \quad
  \inferrule*[lab=\rulelabel{S-NMrgV}]
    { }
    {\SStep{v}{v_1 \nmrg v_2}{v_1 \mrg v_2}}}
\rulerow{
  \inferrule*[lab=\rulelabel{S-Beta}]
    { }
    {\SStep{v}{\clos{v_2}{A}{e}\; v_1}{(v_2 \mrg v_1) \triangleright e}}
  \qquad
  \inferrule*[lab=\rulelabel{S-IdxV}]
    {\lookupV(v_1, n) = v_2}
    {\SStep{v}{v_1.n}{v_2}}
  \qquad
  \inferrule*[lab=\rulelabel{S-SelV}]
    {\selj{v_1}{l}{v_2}}
    {\SStep{v}{v_1.l}{v_2}}}
\rulerow{
  \inferrule*[lab=\rulelabel{S-LetV}]
    { }
    {\SStep{v}{\letb{v_1}{e_2}}{(v \mrg v_1) \triangleright e_2}}
  \qquad
  \inferrule*[lab=\rulelabel{S-OpenV}]
    { }
    {\SStep{v}{\openm{\lrec{l}{v'}}{e_2}}{(v \mrg v') \triangleright e_2}}}
\rulerow{
  \inferrule*[lab=\rulelabel{S-Fun}]
    { }
    {\SStep{v}{\mfunctor{s}{A}{e}}{\mclos{\ambm{s}{v}}{A}{e}}}
  \quad
  \inferrule*[lab=\rulelabel{S-MBeta}]
    { }
    {\SStep{v}{\mapp{\mclos{v_2}{A}{e}}{v_1}}{(v_2 \mrg v_1)
\triangleright e}}}
\rulerow{
  \inferrule*[lab=\rulelabel{S-LinkV}]
    {\selj{v_1}{l}{v_l}}
    {\SStep{v}{\mlink{v_1}{\mclos{v_2}{\rcd{l}{A}}{e}}}
      {v_1 \mrg ((v_2 \mrg \lrec{l}{v_l}) \triangleright e)}}}
\end{rulefigure}
\caption{Small-step semantics of $\sce$.}
\label{fig:sstep}
\compactfloatend
\end{figure*}

\subsubsection*{Environments in reduction.}
Frames $F$ specify left-to-right evaluation, and rule
\ruleref{S-Frame} preserves the ambient environment. Rules \ruleref{S-Env}, \ruleref{S-BoxV}, \ruleref{S-IdxV} and \ruleref{S-SelV} return the environment, the body's value, or the entry found.
Boxes and dependent merges change the environment used for their
right operands. A box $e_1 \triangleright e_2$ replaces it with
the value of $e_1$ (\ruleref{S-BoxR}), whereas a dependent merge
extends it with the left operand's value (\ruleref{S-MrgR}). Rules \ruleref{S-LetV} and \ruleref{S-OpenV} reduce $\kwe{let}$ and $\kwe{open}$ to boxes whose bodies run under the ambient environment extended with the bound value. A non-dependent merge leaves the environment unchanged: frames evaluate both operands under it, and rule \ruleref{S-NMrgV} then turns the two values into a dependent merge.

A lambda captures the ambient
environment in a closure (\ruleref{S-Lam}). Rule \ruleref{S-Beta} then
runs the body in a box under the captured environment extended with the
argument, without substitution. Functors work in the same way
(\ruleref{S-Fun}, \ruleref{S-MBeta}), except that rule \ruleref{S-Fun} captures $\ambm{s}{v}$, which is the empty environment for a sandboxed functor.
Rules \ruleref{S-MrgR} and \ruleref{S-Beta} both extend the environment on the right, so $\lookupV$ also counts from the right end:
{\compactdisplayskips\[
\begin{aligned}
\lookupV(v_1 \mrg v_2, 0) &= v_2 \\[2pt]
\lookupV(v_1 \mrg v_2, n{+}1) &= \lookupV(v_1, n)
\end{aligned}
\]}

\begin{samepage}
For example, take an ambient environment $v$ and the term
$(\lambda A.\;\lambda B.\;e_1 \mrg e_2)\;v_1\;v_2$. Its two beta steps
run the body under $(v \mrg v_1) \mrg v_2$. The merge then adds the
value $v_3$ of $e_1$ for its right operand. So the indices denote the following:
{\compactdisplayskips\[
\begin{array}{@{}l@{\qquad}c@{\qquad}c@{\qquad}c@{}}
 & \text{Environment} & \underline{0} & \underline{1} \\[3pt]
\text{Left operand } e_1 & (v \mrg v_1) \mrg v_2 & v_2 & v_1 \\[3pt]
\text{Right operand } e_2
  & ((v \mrg v_1) \mrg v_2) \mrg v_3 & v_3 & v_2
\end{array}
\]}
The same index thus denotes different entries in the two operands of
a dependent merge. Index $\underline{1}$ is $v_1$ on the left and $v_2$ on the right. This difference is what rule \ruleref{Elab-Link} relies on. In its term $\underline{1} \mrg (\underline{1}\;\lrec{l}{\underline{0}.l})$, the left $\underline{1}$ denotes the provider. On the right, $\underline{1}$ denotes the client and $\underline{0}$ the provider.
\end{samepage}

Rule \ruleref{S-LinkV} is module application followed by a merge: its right operand is the \ruleref{S-MBeta} reduct of the client applied to $\lrec{l}{v_l}$, and the provider is merged on the left. The link evaluates the provider first, then the client. Below, the provider \code{One} supplies \code{x} to a client \code{Inc} that adds $1$:
{\compactdisplayskips\[
\begin{gathered}
\code{One} = \mstruct{\lrec{\code{x}}{1}},
\qquad
\code{Inc} = \mfunctor{\sandbox}{\rcd{\code{x}}{\Int}}{(\underline{\code{x}} + 1)}
\\[4pt]
\SMStep{\unit}{\mlink{\code{One}}{\code{Inc}}}
  {\code{One} \mrg ((\unit \mrg \lrec{\code{x}}{1}) \triangleright (\underline{\code{x}} + 1))}
\longrightarrow^{*} \code{One} \mrg 2
\end{gathered}
\]}
Under $\unit \mrg \lrec{\code{x}}{1}$, the box runs the sandboxed client \code{Inc}, so its body can access only the selected import. The body evaluates to $2$, and the link to $\code{One} \mrg 2$.

The link above selected \code{x} from \code{One}. Label selection $\selj{v}{l}{v'}$ searches both sides of a merge and looks through structures (\ruleref{Sel-MrgL}, \ruleref{Sel-MrgR}, \ruleref{Sel-Str}). When the type of $v$ contains $l$, only one side has $l$, so selection is unique, which makes well-typed evaluation deterministic (Theorem~\ref{thm:determinism}).

Rule \ruleref{S-LinkV} uses only selection, a record, a box and merges, all of which $\lame$ has. Rule \ruleref{Elab-Link} expresses this behavior as a $\lame$ term (Section~\ref{sec:tde}), so $\lame$ needs no separate linking rule.

\subsection{Type Safety}\label{sec:scesafety}

We state the typing properties of $\sce$ using the elaboration
judgment with its core term hidden:
{\compactdisplayskips\[
\STyp{\Gamma}{e}{A} \;\triangleq\; \exists\, \cterm{e'},\, \Elab{\Gamma}{e}{A}{e'}
\]}
An expression has at most one type under a given context.

\begin{theorem}[Inference Uniqueness]\label{thm:inference-uniqueness}
If\/ $\STyp{\Gamma}{e}{A_1}$ and $\STyp{\Gamma}{e}{A_2}$,
then $A_1 = A_2$.
\end{theorem}

A value has the same type under every context.

\begin{lemma}[Value Weakening]\label{lem:val-weaken}
If\/ $\STyp{\Gamma_1}{v}{A}$, then $\STyp{\Gamma_2}{v}{A}$ for every $\Gamma_2$.
\end{lemma}

Indexed lookup and label selection succeed on a well-typed value and return an entry of the type that lookup or containment gives.

\begin{lemma}[Lookup Progress]\label{lem:lookup-prog}
If\/ $\STyp{\varepsilon}{v}{A}$ and $\lookup(A, n) = B$,
then $\lookupV(v, n) = v'$ for some value $v'$ with $\STyp{\varepsilon}{v'}{B}$.
\end{lemma}

\begin{lemma}[Selection Progress]\label{lem:sel-prog}
If\/ $\STyp{\varepsilon}{v}{A}$ and $\lin{l}{B}{A}$,
then $\selj{v}{l}{v'}$ for some value $v'$ with $\STyp{\varepsilon}{v'}{B}$.
\end{lemma}

Boxes and dependent merges change the environment during evaluation,
so we prove progress and preservation for an arbitrary environment $v$
of type $\Gamma$. Because environments are values, the expression judgment types them too.

\begin{theorem}[Progress]\label{thm:progress}
If\/ $\STyp{\varepsilon}{v}{\Gamma}$ and $\STyp{\Gamma}{e}{A}$,
then $e$ is a value or $\SStep{v}{e}{e'}$ for some $e'$.
\end{theorem}

\begin{theorem}[Preservation]\label{thm:preservation}
If\/ $\STyp{\varepsilon}{v}{\Gamma}$, $\STyp{\Gamma}{e}{A}$,
and $\SStep{v}{e}{e'}$, then $\STyp{\Gamma}{e'}{A}$.
\end{theorem}

Although variables are de Bruijn indices, the proofs need no term-substitution or index-shifting lemmas: they use environment lookup and value weakening. For linking, Lemma~\ref{lem:sel-prog} supplies the import value at the required type. The typing rules for records, boxes, and
merges then establish preservation of the linked result. Instantiating both theorems with $v = \unit$ and $\Gamma = \varepsilon$ gives type safety.

\begin{corollary}[Type Safety]\label{cor:type-safety}
If\/ $\STyp{\varepsilon}{e}{A}$, then $e$ is a value or $\SStep{\unit}{e}{e'}$ for some $e'$, and every such $e'$ satisfies $\STyp{\varepsilon}{e'}{A}$.
\end{corollary}
 
\section{Elaboration and Separate Compilation
Correctness}\label{sec:elaboration}
Modules in $\sce$ elaborate to terms of the core calculus
$\lame$~\cite{lambdaeOOPSLA24}. The elaboration of a source link is
exactly the term that the core linker emits for the compiled fragments.
Compiling and linking therefore commute by construction (the
upper square of Figure~\ref{fig:overview-correctness}). Separate
compilation correctness then follows from semantics preservation (the
lower square) and behavior preservation, for static and dynamic links alike. Section~\ref{sec:lame} fixes notation for $\lame$, and
Section~\ref{sec:tde} presents the elaboration.
Section~\ref{sec:separate-compilation} proves correctness and interface
abstraction.

\subsection{Target calculus \texorpdfstring{$\lame$}{lambda-E}}
\label{sec:lame}

The calculus $\lame$ has the expression and type forms shown in black
in Figure~\ref{fig:sce-syntax}. \citet{lambdaeOOPSLA24} give the typing
and semantics of $\lame$.
We write $\svdash$ for source judgments and $\cvdash$ for core
judgments. Where both calculi appear, $u$ ranges over core values and
$v$ over source values.

\subsubsection*{Typing and Semantics.}
The typing judgment $\HasType{\Gamma}{e}{A}$ of $\lame$ uses a type
as its context $\Gamma$, as in $\sce$.
$\lame$ is type safe~\cite{lambdaeOOPSLA24}, and our mechanization
reproves progress and preservation for $\lame$.

\subsection{Type-Directed Elaboration}\label{sec:tde}
Elaboration turns a structure into its body, a functor into a function,
and a link into a core term built from applications, selection, and merge.
The target therefore needs no module constructs.
Figures~\ref{fig:elaboration-core} and~\ref{fig:elaboration-module}
pair each typing rule with its core term, and this subsection explains
those terms.

\subsubsection*{Type Translation.}
The translation $\elab{\cdot}$ maps source types to $\lame$ types. It
is homomorphic on the constructors shared with $\lame$, removes
interface wrappers, and turns functor types into arrows:
{\compactdisplayskips\[
\elab{\Intf\; A} \;=\; \elab{A}
\qquad\qquad
\elab{A \marr B} \;=\; \elab{A} \to \elab{B}
\]}
Since contexts are types, the same translation applies to the typing
context $\Gamma$.

\subsubsection*{Core Expressions.}
Elaboration preserves the expression forms shared with $\lame$ and
recursively elaborates their subterms. Rules \ruleref{Elab-Lam} and
\ruleref{Elab-Clos} also translate the parameter annotation.

\subsubsection*{Derived Expressions.}
Let bindings, open expressions, and non-dependent merges elaborate to
core environment operations. The examples
evaluate under the ambient environment $\lrec{\code{x}}{7}$:
{\compactdisplayskips\[
\begin{array}{@{}l|l|l@{}}
\text{Source expression} & \text{Elaborated term}
  & \text{Value} \\[2pt]
\hline
\rule{0pt}{2.8ex}\letb{\lrec{\code{y}}{1}}{\underline{\code{y}} + \underline{\code{x}}}
  & (\lrec{\code{y}}{1} \mrg (\underline{\code{y}} + \underline{\code{x}})).0
  & 8 \\[2pt]
\openm{\lrec{\code{m}}{\lrec{\code{y}}{1}}}{\underline{\code{y}} + \underline{\code{x}}}
  & (\lambda \rcd{\code{y}}{\Int}.\; \underline{\code{y}} + \underline{\code{x}})\;
      (\lrec{\code{m}}{\lrec{\code{y}}{1}}.\code{m})
  & 8 \\[2pt]
\lrec{\code{x}}{1} \nmrg \lrec{\code{z}}{\underline{\code{x}} + 1}
  & \bigl(\query \mrg ((\underline{0} \triangleright \lrec{\code{x}}{1})
      \mrg (\underline{1} \triangleright
        \lrec{\code{z}}{\underline{\code{x}} + 1}))\bigr).0
  & \lrec{\code{x}}{1} \mrg \lrec{\code{z}}{8}
\end{array}
\]}

A let binding uses a dependent merge to extend the body's environment
with the whole bound value, then projects the body's result
(\ruleref{Elab-Let}). An open expression instead extends the environment
with only the field value of its operand (\ruleref{Elab-Open}). The
operand's synthesized type $\rcd{l}{A}$
supplies the selection label $l$ and the lambda annotation $\elab{A}$:
\code{m} and $\rcd{\code{y}}{\Int}$ in the second example. Only the open
encoding uses type information.

\looseness=-1 In a non-dependent merge, the outer $\query$ saves the ambient
environment, which boxes restore for each operand
(\ruleref{Elab-NMrg}). In the third example, the right operand
sees the ambient $\code{x} = 7$, so $\code{z} = 8$.

\subsubsection*{Modules.}
Elaboration removes structure wrappers (\ruleref{Elab-Str}) and maps
module closures and applications to their core counterparts
(\ruleref{Elab-MClos}, \ruleref{Elab-MApp}).
An open functor elaborates to a lambda abstraction that captures the
ambient environment (\rulerefx{Elab-Fun-op}{Elab-Fun$_\open$}).
A sandboxed functor elaborates to the boxed abstraction
$\unit \triangleright (\lambda \elab{A}.\; e')$
(\rulerefx{Elab-Fun-sb}{Elab-Fun$_\sandbox$}). The box makes the
abstraction capture the empty environment.
Rule \rulerefx{Elab-Fun-sb}{Elab-Fun$_\sandbox$} ignores the ambient
context, so a sandboxed functor elaborates the same way wherever it
occurs. A sandboxed client therefore compiles once and links in any
context:

\begin{lemma}[Context Independence]\label{lem:context-independence}
If\/ $\Elab{\Gamma_1}{\mfunctor{\sandbox}{A}{e}}{C}{e'}$, then
{\compactdisplayskips\[
\Elab{\Gamma_2}{\mfunctor{\sandbox}{A}{e}}{C}{e'}
\qquad\text{for every }\Gamma_2.
\]}
\end{lemma}

\begin{proof}
Rule \rulerefx{Elab-Fun-sb}{Elab-Fun$_\sandbox$} checks the body under
$\varepsilon \mathbin{\&} A$, independently of the ambient context.
\end{proof}

\subsubsection*{Linking.}
A link $\mlink{e_1}{e_2}$, with operand types $\Gamma_1$ and
$\rcd{l}{A} \marr B$, elaborates by rule \ruleref{Elab-Link} to
the following term. The named form identifies the provider $p$ and
client $f$:
\begin{equation}\label{eq:linkc}
\begin{array}{@{}l|l@{}}
\text{Named}
  & (\lambda (p : \elab{\Gamma_1}).\;
      \lambda (f : \rcd{l}{\elab{A}} \to \elab{B}).\;
      p \mrg (f\; \lrec{l}{p.l}))\; e_1'\; e_2' \\[3pt]
\text{Nameless}
  & (\lambda \elab{\Gamma_1}.\; \lambda (\rcd{l}{\elab{A}} \to \elab{B}).\;
      \underline{1} \mrg (\underline{1}\; \lrec{l}{\underline{0}.l}))\;
      e_1'\; e_2'
\end{array}
\end{equation}
We write
$\linkc(\elab{\Gamma_1},\rcd{l}{\elab{A}},\elab{B},e_1',e_2')$
for this term. It generalizes the term $e_\ell$ that links
$\cterm{m'}$ and $\cterm{n'}$ in the overview to any import label and interfaces.

Across all rules, elaboration preserves types
(Theorem~\ref{thm:type-preservation}) and produces a unique core term
(Theorem~\ref{thm:elaboration-uniqueness}). With
Theorem~\ref{thm:inference-uniqueness}, a well-typed expression and its
context therefore determine the expression's type and elaboration.
The elaboration of a link is thus well defined.

\begin{theorem}[Type Preservation]\label{thm:type-preservation}
If\/ $\Elab{\Gamma}{e}{A}{e'}$ then $\HasType{\elab{\Gamma}}{e'}{\elab{A}}$.
\end{theorem}

\begin{theorem}[Elaboration Uniqueness]\label{thm:elaboration-uniqueness}
If\/ $\Elab{\Gamma}{e}{A_1}{e_1'}$ and $\Elab{\Gamma}{e}{A_2}{e_2'}$
then $e_1' = e_2'$.
\end{theorem}

\subsection{Separate Compilation Correctness}
\label{sec:separate-compilation}

A sandboxed client is compiled against only the \emph{interface} of
its import, and links with any provider whose type contains it. We derive separate compilation correctness from semantics and behavior
preservation, and show that the client's result depends on its provider
only through the import.

\subsubsection*{Big-Step Semantics.}
We use big-step evaluation to state semantics preservation. The
judgment $\SBig{v}{e}{v'}$ states that $e$ evaluates to the value $v'$
under the ambient environment $v$. Its rules evaluate subexpressions
under the same environments as the small-step rules, and
Appendix~\ref{app:sce-bigstep} gives the rules in full. The rule for a link
mirrors rule \ruleref{S-LinkV}:
\begin{mathpar}
  \inferrule*[lab=\rulelabel{Eval-Link}]
    {\SBig{v}{e_1}{v_1} \\
     \SBig{v}{e_2}{\mclos{v_2}{\rcd{l}{A}}{e}} \\
     \selj{v_1}{l}{v_l} \\
     \SBig{v_2 \mrg \lrec{l}{v_l}}{e}{v_3}}
    {\SBig{v}{\mlink{e_1}{e_2}}{v_1 \mrg v_3}}
\end{mathpar}
Big-step evaluation agrees with small-step reduction to a value. The
results on terminating runs below use big-step evaluation, and
Theorem~\ref{thm:equiv} carries them to small-step reduction:

\begin{theorem}[Big-Step/Small-Step Equivalence]\label{thm:equiv}
$\SBig{v}{e}{v'}$ if and only if\/ $\SMStep{v}{e}{v'}$.
\end{theorem}

For $\lame$, we write $\EBig{u}{e}{u'}$ for big-step evaluation, which
the mechanization also proves equivalent to small-step reduction.

\subsubsection*{Semantics Preservation.}
Elaboration preserves evaluation: a terminating source expression and
its elaboration produce values related by elaboration. The proof uses
three facts: source
values elaborate to core values, and indexed lookup and label selection
commute with elaboration. We use the same notation for the corresponding
core operations.

\begin{lemma}[Value Elaboration]\label{lem:elab-value}
If\/ $\Elab{\Gamma}{v}{A}{e'}$, then $e'$ is a value.
\end{lemma}

\begin{lemma}[Lookup Preservation]\label{lem:lookup-preservation}
If\/ $\Elab{\varepsilon}{v}{A}{u}$, $\lookup(A, n) = B$, and
$\lookupV(v, n) = v'$, then
$\exists\, u',\, \lookupV(u, n) = u'$ and
$\Elab{\varepsilon}{v'}{B}{u'}$.
\end{lemma}

\begin{lemma}[Selection Preservation]\label{lem:sel-preservation}
If\/ $\Elab{\varepsilon}{v}{A}{u}$, $\lin{l}{B}{A}$, and
$\selj{v}{l}{v'}$, then
$\exists\, u',\, \cselj{u}{l}{u'}$ and
$\Elab{\varepsilon}{v'}{B}{u'}$.
\end{lemma}

\begin{theorem}[Semantics Preservation]\label{thm:semantics-preservation}
If\/ $\Elab{\Gamma}{e}{A}{e'}$,
$\Elab{\varepsilon}{v}{\Gamma}{u}$, and
$\SBig{v}{e}{v'}$,
then $\exists\, u',\, \EBig{u}{e'}{u'}$ and
$\Elab{\varepsilon}{v'}{A}{u'}$.
\end{theorem}

\begin{proof}[Proof Sketch]
By induction on the source evaluation, using
Lemmas~\ref{lem:elab-value}--\ref{lem:sel-preservation} for queries,
indexed projections, and field selection. In the linking case, the
induction hypotheses and Lemma~\ref{lem:sel-preservation} give the core provider, client, and
export, and the term~\eqref{eq:linkc} runs the client body as
rule \ruleref{Eval-Link} does.
\end{proof}

Theorem~\ref{thm:semantics-preservation} runs from source to core and speaks only of
terminating runs. Theorem~\ref{thm:behavior-preservation} adds the converse and
divergence. Big-step evaluation cannot express a run that never ends,
so we write $\SDiv{v}{e}$ when $e$ has an infinite small-step reduction under
$v$, and likewise $\EDiv{u}{e'}$ for $\lame$.

\begin{theorem}[Behavior Preservation]\label{thm:behavior-preservation}
If\/ $\Elab{\Gamma}{e}{A}{e'}$ and $\Elab{\varepsilon}{v}{\Gamma}{u}$, then
\begingroup\listisep=0pt\begin{itemize}\compactitems
\item[(i)] $\EBig{u}{e'}{u'}$ implies $\SBig{v}{e}{v'}$ and
$\Elab{\varepsilon}{v'}{A}{u'}$ for some $v'$; and
\item[(ii)] $\SDiv{v}{e}$ if and only if $\EDiv{u}{e'}$.
\end{itemize}\endgroup
\end{theorem}

\begin{proof}[Proof Sketch]
An auxiliary simulation relation matches each source step with
zero or more core steps. Zero-step matches strictly decrease a
natural-number measure on source terms, so infinite source reductions
yield infinite core reductions. Termination and divergence reflection
then follow from source type safety, semantics preservation, and
core determinism.
\end{proof}

Fixpoints or iso-recursive types, added in
Section~\ref{sec:extensions}, let terms diverge. Like every theorem in the paper,
Theorem~\ref{thm:behavior-preservation} covers that extended language.

Taking $\Gamma = \varepsilon$ and $v = u = \unit$ in
Theorems~\ref{thm:semantics-preservation} and~\ref{thm:behavior-preservation}
gives correctness for whole programs.

\begin{corollary}[Whole-Program Correctness]\label{cor:whole-program}
If\/ $\Elab{\varepsilon}{e}{A}{e'}$, then $e$ and $e'$ have the same
behavior under $\unit$:
\begingroup\listisep=0pt\begin{itemize}\compactitems
\item[(a)] $\SBig{\unit}{e}{v'}$ implies $\EBig{\unit}{e'}{u'}$ and
$\Elab{\varepsilon}{v'}{A}{u'}$ for some $u'$;
\item[(b)] $\EBig{\unit}{e'}{u'}$ implies $\SBig{\unit}{e}{v'}$ and
$\Elab{\varepsilon}{v'}{A}{u'}$ for some $v'$; and
\item[(c)] $\SDiv{\unit}{e}$ if and only if $\EDiv{\unit}{e'}$.
\end{itemize}\endgroup
\end{corollary}

\subsubsection*{Core Linking.}
Corollary~\ref{cor:whole-program} covers a program compiled as a whole.
The \emph{core linker} instead links separately compiled fragments. It receives
two core fragments and their interfaces, not their sources. We model
it by the judgment
{\compactdisplayskips\[
  \clink{I_1}{e_1}{\rcd{l}{I_l} \to I_2}{e_2}{e_\ell}
\]}
The judgment links a provider $e_1$ of interface $I_1$ and a
client $e_2$ into the term $e_\ell$. Here the client imports
a field $l$ of type $I_l$ and exports $I_2$. The interfaces $I_1$,
$I_l$ and $I_2$ play the roles of $\elab{\Gamma_1}$, $\elab{A}$ and
$\elab{B}$ in term~\eqref{eq:linkc}.
Rule \ruleref{CL-Link} checks containment and emits the term of
rule \ruleref{Elab-Link}:
\begin{mathpar}
  \inferrule*[lab=\rulelabel{CL-Link}]
    {\linc{l}{I_l}{I_1}}
    {\clink{I_1}{e_1}{\rcd{l}{I_l} \to I_2}{e_2}
      {\linkc(I_1,\rcd{l}{I_l},I_2,e_1,e_2)}}
\end{mathpar}
The judgment uses only the fragments and their interfaces, so it has
no context index and does not check the fragments' types.
Theorem~\ref{thm:core-link-typed} assumes their typing under any core
context $\Gamma_c$.

\begin{theorem}[Link Typing]\label{thm:core-link-typed}
If\/ $\clink{I_1}{e_1}{\rcd{l}{I_l} \to I_2}{e_2}{e_\ell}$,
$\HasType{\Gamma_c}{e_1}{I_1}$, and
$\HasType{\Gamma_c}{e_2}{\rcd{l}{I_l} \to I_2}$,
then $\HasType{\Gamma_c}{e_\ell}{I_1 \mathbin{\&} I_2}$.
\end{theorem}

The result keeps both modules' exports, so it can be the provider of
a further link.

\subsubsection*{Reusing a Compiled Client.}
Rule \ruleref{CL-Link} lets one compiled client serve providers it was
never compiled against. Take $\cterm{n'}$, the compiled \code{Double} of
Section~\ref{sec:linking-inside}, whose interface is
$\rcd{\code{op}}{\Int \to \Int \to \Int} \to \code{T}$ with
$\code{T} = \rcd{\code{double}}{\Int \to \Int}$, and a provider
compiled with another interface:
{\compactdisplayskips\[
e_1 \;=\; \lrec{\code{unit}}{1} \mrg
  \lrec{\code{op}}{\lambda \Int.\;\lambda \Int.\;
    \underline{1} \times \underline{0}}
\qquad\qquad
I_1 \;=\; \rcd{\code{unit}}{\Int} \mathbin{\&}
  \rcd{\code{op}}{\Int \to \Int \to \Int}
\]}
Since $\linc{\code{op}}{\Int \to \Int \to \Int}{I_1}$ holds, rule
\ruleref{CL-Link} derives
{\compactdisplayskips\[
\clink{I_1}{e_1}{\rcd{\code{op}}{\Int \to \Int \to \Int} \to \code{T}}{\cterm{n'}}{e_\ell}
\]}
Here \code{op} multiplies, so $e_\ell.\code{double}\;5$ evaluates to $25$.
Type equality is structural, so $I_1$ differs from the interface \code{P} of
the compiled \code{Add}, which lists \code{op} first. The client $\cterm{n'}$ is still reused
unchanged, because rule \ruleref{CL-Link} checks only that $I_1$
contains \code{op}.

\subsubsection*{Separate Compilation.}
Separate compilation correctness is a corollary of semantics and
behavior preservation, because the core linker emits the elaboration of the
source link. The provider and the client may depend on the ambient
context $\Gamma$, for example when a function receives its provider as
an argument. The linked core term then runs under the elaboration of
the source environment.

In the corollaries, the provider's interface $I_1$, the import type
$I_l$ and the client's exports $I_2$ are source types. The core linker
receives their translations $\elab{I_1}$, $\elab{I_l}$ and $\elab{I_2}$.

\begin{samepage}
\begin{corollary}[Dynamic Separate Compilation]
\label{thm:separate-compilation}
If
{\compactdisplayskips\[
\begin{array}{@{}l@{\qquad}l@{}}
\text{(1)}\ \Elab{\Gamma}{e_1}{I_1}{e_1'} \wedge \lin{l}{I_l}{I_1} &
\text{(2)}\ \Elab{\Gamma}{e_2}{\rcd{l}{I_l} \marr I_2}{e_2'}\\
\text{(3)}\ \clink{\elab{I_1}}{e_1'}{\rcd{l}{\elab{I_l}} \to \elab{I_2}}{e_2'}{e_\ell} &
\text{(4)}\ \Elab{\varepsilon}{v}{\Gamma}{u}
\end{array}
\]}
Then $\mlink{e_1}{e_2}$ under $v$ and $e_\ell$ under $u$ have the same
behavior:
\begingroup\listisep=0pt\begin{itemize}\compactitems
\item[(a)] $\SBig{v}{\mlink{e_1}{e_2}}{v'}$ implies $\EBig{u}{e_\ell}{u'}$ and
$\Elab{\varepsilon}{v'}{I_1 \mathbin{\&} I_2}{u'}$ for some $u'$;
\item[(b)] $\EBig{u}{e_\ell}{u'}$ implies $\SBig{v}{\mlink{e_1}{e_2}}{v'}$ and
$\Elab{\varepsilon}{v'}{I_1 \mathbin{\&} I_2}{u'}$ for some $v'$; and
\item[(c)] $\SDiv{v}{\mlink{e_1}{e_2}}$ if and only if $\EDiv{u}{e_\ell}$.
\end{itemize}\endgroup
\end{corollary}
\end{samepage}

\begin{proof}[Proof Sketch]
With (1)--(3) and rules \ruleref{CL-Link} and \ruleref{Elab-Link}, we
have $\Elab{\Gamma}{\mlink{e_1}{e_2}}{I_1 \mathbin{\&} I_2}{e_\ell}$.
Theorem~\ref{thm:semantics-preservation}, applied to this elaboration
with (4), yields (a), and Theorem~\ref{thm:behavior-preservation} yields
(b) and (c).
\end{proof}

A sandboxed client is compiled once, under $\varepsilon$, before any
provider or scope is chosen: by Lemma~\ref{lem:context-independence}, its
compiled term satisfies (2) for every $\Gamma$. An open client's compiled
term may refer to $\Gamma$, so it is valid only under the context
$\Gamma$ in which it was compiled. The corollary also covers such
clients, each in its own context, as well as clients that are variables
bound to loaded modules. Loading itself lies outside the calculus: through
premise (4), the corollary assumes that a loaded module's value
elaborates to its compiled form. Static linking instead compiles
both fragments under $\varepsilon$ and runs them under the empty
environment.

\begin{corollary}[Static Separate Compilation]\label{thm:static-sc}
Assume (1)--(3) of Corollary~\ref{thm:separate-compilation} with
$\Gamma = \varepsilon$. Then (a)--(c) hold with $v = u = \unit$.
\end{corollary}

\begin{proof}
Apply Corollary~\ref{thm:separate-compilation} with $v = u = \unit$.
\end{proof}

In words, the core linker's output evaluates to the elaboration of
the source link's result exactly when the source link terminates, and
diverges exactly when the source link does. Both corollaries give this
for a client compiled once and a separately compiled provider.

The corollaries cover a single link. An \textsf{SCE-Lang} program is a sequence
of files, each linked against the earlier ones, or merged with them if
it has no imports.
Theorem~\ref{thm:unit-chain} extends Corollary~\ref{thm:static-sc} to such a sequence
of binary links and merges with import-free files
(Appendix~\ref{app:units-linking}).

\subsubsection*{Interface Abstraction.}
\looseness=-1 A new provider preserves the client's result if the selected
import value is unchanged. We fix the client, ambient environment, and
import value, with its captured environments.

\begin{theorem}[Interface Abstraction]\label{thm:interface-abstraction}
Let $p_1$ and $p_2$ be values. Suppose
$\STyp{\Gamma}{e}{\rcd{l}{A} \marr B}$,
$\STyp{\varepsilon}{v}{\Gamma}$, and, for $i = 1,2$,
$\STyp{\varepsilon}{p_i}{P_i}$, $\lin{l}{A}{P_i}$, and
$\selj{p_i}{l}{v_l}$. Then, for every value $w$,
{\compactdisplayskips\[
\SBig{v}{\mlink{p_1}{e}}{p_1 \mrg w}
\quad\Longleftrightarrow\quad
\SBig{v}{\mlink{p_2}{e}}{p_2 \mrg w}
\]}
and $\SDiv{v}{\mlink{p_1}{e}}$ if and only if $\SDiv{v}{\mlink{p_2}{e}}$.
\end{theorem}

\begin{proof}[Proof Sketch]
Typing makes selection of $l$ unique, so rule \ruleref{Eval-Link}
supplies the same record $\lrec{l}{v_l}$ to the client in both links.
Reusing the client and body derivations proves each direction. A
well-typed link diverges exactly when it does not terminate, which gives
the divergence part.
\end{proof}

\begin{figure}[t]
\compactcaption
\centering
\small
\begin{tabular}{@{}p{0.23\linewidth}@{\quad}p{0.73\linewidth}@{}}
\toprule
\textbf{Property} & \textbf{Description} \\
\midrule
\raggedright Context independence\par
(Lemma~\ref{lem:context-independence})
& \raggedright A sandboxed functor
$f = \mfunctor{\sandbox}{A}{e}$ has the same type and elaborates
to the same core term in every context.
\par\smallskip
$\Elab{\Gamma_1}{f}{C}{e'} \;\Longrightarrow\;
 \Elab{\Gamma_2}{f}{C}{e'}$,
for all $\Gamma_1,\Gamma_2$.
\tabularnewline[3pt]

\raggedright Interface abstraction\par
(Theorem~\ref{thm:interface-abstraction})
& \raggedright Fix a client $e$ of type $\rcd{l}{A}\marr B$ under
$\Gamma$, and an environment $v$ of type $\Gamma$. Provider values
$p_1,p_2$ that contain $\rcd{l}{A}$ in their types and select the same
import value give
\par\smallskip
$\SBig{v}{\mlink{p_1}{e}}{p_1\mrg w}
  \;\Longleftrightarrow\;
  \SBig{v}{\mlink{p_2}{e}}{p_2\mrg w}$.
\tabularnewline[3pt]
\raggedright Separate compilation\par
(Corollary~\ref{thm:separate-compilation};
Corollary~\ref{thm:static-sc})
& \raggedright Linking independently compiled modules preserves the
result of their source link. For
$\STyp{\Gamma}{\mlink{e_1}{e_2}}{I_1\mathbin{\&}I_2}$,
let $e_\ell$ link their separately elaborated core terms.
If $\Elab{\varepsilon}{v}{\Gamma}{u}$, then
\par\smallskip
$\SBig{v}{\mlink{e_1}{e_2}}{v'}\Longrightarrow\exists\,u',\,
\EBig{u}{e_\ell}{u'}\land
\Elab{\varepsilon}{v'}{I_1\mathbin{\&}I_2}{u'}$.
\tabularnewline
\bottomrule
\end{tabular}
\caption{Properties of linking and separate compilation.}
\label{fig:linking-properties}
\compactfloatend
\end{figure}

Figure~\ref{fig:linking-properties} summarizes context independence,
interface abstraction, and separate compilation correctness.
Its entries omit divergence, and the separate compilation
entry shows only clause~(a).

\subsubsection*{Sandbox and Link Confinement.}
A link never passes its own ambient environment to the client. The
client runs under the environment it captured, extended with the
selected import. A sandboxed client captures the empty environment
(\ruleref{S-Fun}), even inside a function, as the box in its elaboration
(\rulerefx{Elab-Fun-sb}{Elab-Fun$_\sandbox$}) does.
Its body therefore sees only that import: no binding in scope where the
client is written or linked, and no other export of the provider.
Since typing makes every selection unique, source evaluation is deterministic, and the agreement of results follows from determinism, both for a
sandboxed functor applied directly and for links. For divergence, both runs take the same first
steps, then run the same body under the same environment, so they diverge together.

\begin{theorem}[Determinism of Evaluation]\label{thm:determinism}
If\/ $\STyp{\Gamma}{e}{A}$, $\STyp{\varepsilon}{v}{\Gamma}$,
$\SBig{v}{e}{v_1}$, and $\SBig{v}{e}{v_2}$, then $v_1 = v_2$.
\end{theorem}

\begin{theorem}[Sandbox Confinement]\label{thm:confinement}
Let $\STyp{\varepsilon \mathbin{\&} A}{e}{B}$,
$\STyp{\varepsilon}{v_l}{A}$, and $f = \mfunctor{\sandbox}{A}{e}$.
For values $\rho_1,\rho_2$, if $\SBig{\rho_i}{\mapp{f}{v_l}}{v_i}$ for $i=1,2$, then $v_1=v_2$. Moreover,
$\SDiv{\rho_1}{\mapp{f}{v_l}}$ if and only if $\SDiv{\rho_2}{\mapp{f}{v_l}}$.
\end{theorem}

\begin{theorem}[Link Confinement]\label{thm:link-confinement}
Let $c = \mclos{\unit}{\rcd{l}{A}}{e}$ with
$\STyp{\varepsilon}{c}{\rcd{l}{A}\marr B}$, and
$\STyp{\varepsilon}{v_l}{A}$. For $i=1,2$, let $p_i$ be a value with
$\selj{p_i}{l}{v'}$ exactly when $v'=v_l$. For values $\rho_1,\rho_2$, if
$\SBig{\rho_i}{\mlink{p_i}{c}}{p_i\mrg w_i}$ for $i=1,2$, then
$w_1=w_2$. Moreover, $\SDiv{\rho_1}{\mlink{p_1}{c}}$ if and only if
$\SDiv{\rho_2}{\mlink{p_2}{c}}$.
\end{theorem}

The client's exports $w_i$ thus depend neither on the ambient
environments $\rho_i$ nor on the rest of the providers.
The closure $c$ captures the empty environment, which excludes a client
that captured a nonempty one, such as an inline open functor. A sandboxed client evaluates to a closure over the empty
environment wherever it is written (Section~\ref{sec:linking-inside}).
For terminating runs, the client may be any expression that evaluates
only to this closure, such as a variable bound to a loaded module.
 \section{Extensions, Implementation and Discussion}
\label{sec:design-implementation}

We extend the calculus and its metatheory
(Section~\ref{sec:extensions}), present the WebAssembly compiler
(Section~\ref{sec:implementation}), and discuss limitations and
further extensions (Section~\ref{sec:discussion}).

\subsection{Extensions}
\label{sec:extensions}

We add unions, recursive types, and fixpoints to both calculi, and
multi-import linking to $\sce$ (Figure~\ref{fig:ext-syntax}).
Every theorem so far extends to these constructs,
and Lean checks the new proof cases. 

\subsubsection*{Unions and Recursion.}
Union types let the loader of Section~\ref{sec:programming-with-linking}
return either a plugin or an error. Matching the \code{Plugin} case
binds the loaded plugin at the functor type required for linking. Iso-recursive
types describe recursive data, and fixpoints define recursive functions.
These constructs elaborate directly to their counterparts in $\lame$;
Appendix~\ref{app:ext-lang} gives the rules.

\begin{figure}[t]
\compactcaption
\centering
{\noindent\formalrow{
\begin{array}{@{}l@{\qquad}rcl@{}}
\syntaxcat{Types} & A, B, \Gamma & ::= & \cdots \mid A \vee B \mid \alpha \mid \mu\alpha.\,A \\
\syntaxcat{Expressions} & e & ::= & \cdots \mid \injl{B}{e} \mid \injr{A}{e} \mid \scase{e}{e_1}{e_2} \\
  & & \mid & \fixlam{A}{B}{e} \mid \tfold{\mu\alpha.A}{e} \mid \tunfold{e} \\
\syntaxcat{Import interfaces} & D & ::= & \rcd{l}{A} \mid D \mathbin{\&} \rcd{l}{A} \\
\syntaxcat{Modules} & m & ::= & \cdots \mid \mlinkn{e_1}{e_2}
\end{array}
}\par}
\caption{Unions and recursion extend both calculi, and multi-import linking extends $\sce$.}
\label{fig:ext-syntax}
\compactfloatend
\end{figure}

\subsubsection*{Multi-Import Linking.}
Multi-import linking $\mlinkn{e_1}{e_2}$ selects a unique provider field
of the required type for each import in
$D = \rcd{l_1}{A_1} \mathbin{\&} \cdots \mathbin{\&} \rcd{l_n}{A_n}$.
Elaboration replaces the single-field record in term~\eqref{eq:linkc}
with $\lrec{l_1}{p.l_1} \mrg \cdots \mrg \lrec{l_n}{p.l_n}$,
where $p$ names the provider.
Core linking constructs the same term from separately compiled modules
(\ruleref{CL-LinkN}): source and core links terminate with results related
by elaboration or diverge together, for both static and dynamic links.
For a fixed well-typed client and ambient environment, links with typed
provider values supplying the same import package yield the same client
exports and diverge together.
Separate compilation correctness also covers files combined in dependency
order by linking against accumulated exports or merging import-free files.
Rules and proofs appear in Appendix~\ref{app:ext-nary}
(Theorems~\ref{thm:static-sc-n} and~\ref{thm:interface-abstraction-n})
and Appendix~\ref{app:units-linking}
(Theorems~\ref{thm:unit-chain} and~\ref{thm:unit-chain-n}).

\FloatBarrier
\subsection{Implementation}
\label{sec:implementation}

\textsf{SCE-Lang} implements $\sce$ in about 4,500 lines of OCaml.
The compiler parses and desugars source programs into $\sce$ terms,
then typechecks and elaborates them into $\lame$. It provides
interpreters for both calculi and a WasmGC backend, and implements
every extension above.

\subsubsection*{Units and Linking.}
\textsf{SCE-Lang} represents an importing file as a sandboxed
functor over its imports, and an import-free file as a structure
boxed under the empty environment.
Each unit elaborates under the empty context to a core term, stored
with the interfaces the loader checks. The linker chains
units, saving each result as a unit for staged linking.
Theorem~\ref{thm:unit-chain-n} proves chains correct.

\subsubsection*{Compilation to WebAssembly.}
The WasmGC backend compiles these units without a dedicated linking
instruction. Figure~\ref{fig:wasm} shows the compilation of $\lame$'s
environment operations, omitting tags and casts. At run time, a local
$\rho$ holds an environment reference, and \wi{struct.new} allocates
garbage-collected pairs and closures. The merge code pairs $\rho$ with
the value $x$ of $e_1$ to evaluate $e_2$, then pairs $x$ with the result.
The core linking term~\eqref{eq:linkc} needs only these operations,
index lookup, records, and field selection.
For multiple files, the backend emits a WebAssembly link module that
imports unit entry points and links their results using the core
construction of Theorem~\ref{thm:unit-chain-n}. Saved linked units can
likewise be compiled and linked again. Appendix~\ref{app:wasm} gives
the full rules; Appendix~\ref{app:linking-example} traces the compilation
of \code{Add} and \code{Double}.

\begin{figure}[t]
\compactcaption
\centering
\small
\setlength{\wasmwidth}{0.75\linewidth}
\begin{tabular}{@{}l@{\;\;$=$\;\;}l@{}}
$\comp{\query}{\rho}$ &
  \wasmblock{\wi{local.get} $\rho$} \\[2pt]
$\comp{e_1 \triangleright e_2}{\rho}$ &
  \wasmblock{$\comp{e_1}{\rho}$ \ws \wi{local.set} $\rho'$ \ws
    $\comp{e_2}{\rho'}$} \\[2pt]
$\comp{e_1 \mrg e_2}{\rho}$ &
  \wasmblock{$\comp{e_1}{\rho}$ \ws \wi{local.set} $x$ \\
    \wi{local.get} $\rho$ \ws \wi{local.get} $x$ \ws
      \wi{struct.new} \wty{Pair} \ws \wi{local.set} $\rho'$
      \wnote{$\rho' = (\rho, x)$} \\
    \wi{local.get} $x$ \ws $\comp{e_2}{\rho'}$ \ws
      \wi{struct.new} \wty{Pair}} \\[2pt]
$\comp{\lambda A.\;e}{\rho}$ &
  \wasmblock{\wi{ref.func} \wty{f} \ws \wi{local.get} $\rho$ \ws
      \wi{struct.new} \wty{Clos} \\
    \textrm{where the lifted function} \wty{f}$(\code{self}, x)$
      \textrm{is} \\
    \quad \wi{local.get} $\code{self}$ \ws
      \wi{struct.get} \wty{Clos} \wfield{env} \ws \wi{local.get} $x$ \\
    \quad \wi{struct.new} \wty{Pair} \ws \wi{local.set} $\rho'$
      \wnote{$\rho' = (\code{self}.\code{env}, x)$} \\
    \quad $\comp{e}{\rho'}$} \\[2pt]
$\comp{e_1\;e_2}{\rho}$ &
  \wasmblock{$\comp{e_1}{\rho}$ \ws \wi{local.tee} $c$ \ws $\comp{e_2}{\rho}$ \\
    \wi{local.get} $c$ \ws \wi{struct.get} \wty{Clos} \wfield{fn} \ws
      \wi{call\_ref} \wty{Fn}}
\end{tabular}
\caption{Compiling the environment operations of $\lame$ to WasmGC.}
\label{fig:wasm}
\compactfloatend
\end{figure}

\subsection{Discussion and Limitations}
\label{sec:discussion}

The calculus is simply typed: signatures describe value exports, without
ML-style abstract type components, sharing constraints, generativity,
or polymorphism.
These features would let separately compiled units hide representation
choices and expose reusable interfaces over abstract types.

Subtyping offers a concrete next step. We could add an interface
restriction $\cterm{e : A}$: when the type of $e$ is a subtype of $A$,
a cast restricts its value to $A$, discarding components outside that
interface. This suggests the following module-application rule:
{\compactdisplayskips\[
\inferrule*[right=\textsc{App-Sub}]
  {\Elab{\Gamma}{e_f}{D \marr B}{e_f'} \\
   \Elab{\Gamma}{e_p}{P}{e_p'} \qquad \elab{P} <: \elab{D}}
  {\Elab{\Gamma}{\mapp{e_f}{e_p}}{B}
    {e_f'\,(e_p' : \elab{D})}}
\]}
Here the inserted restriction supplies the functor's import
interface. At the level of interfaces, this would allow applying
\code{Double} directly to \code{Add}: restriction keeps only
\code{op}. Application would thus perform import matching
but return only the client's result: a form of
\emph{linking as application}.
Disjointness at merges and context extensions would keep casts unambiguous.
Such a discipline could draw on compositional programming~\cite{10.1145/3460228}
and type-directed operational semantics~\cite{huang_et_al:LIPIcs.ECOOP.2020.26}.

Sealing and polymorphism call for further extensions. One direction is
to enrich the target with polymorphism and higher-order type operators,
as in System $F_\omega$, while retaining first-class environments.
We could then investigate an elaboration in the style of F-ing
modules~\cite{FingModules2014}, with the aim of supporting ML-style
modules alongside \emph{first-class} linking and environments.
 \section{Related Work}\label{sec:related}

\paragraph{Program fragments and separate compilation.}
\citeauthor{Cardelli97}'s linksets (Section~\ref{sec:overview-existing}) formalize simply typed separate compilation, preserving typing under combination and linking~\cite{Cardelli97}. SMLSC brings linksets to Standard ML~\cite{SwaseyMCH2006}. \citeauthor{SewellPOPL01} proves that two ML-style structures build in sequence exactly when the second builds as a functor on the first~\cite{SewellPOPL01}. Manifest types~\cite{10.1145/174675.176926}, Backpack~\cite{10.1145/2535838.2535884} and MTAL~\cite{GlewMorrisettPOPL99} check units, packages and object files against imports. All link outside the language; none proves that linked code matches the merged sources. Corollary~\ref{thm:static-sc} proves this property.

\paragraph{Linking as a language construct.}
In Units, a \code{compound} expression links first-class units even inside functions~\cite{FlattFelleisen1998}. \citeauthor{OwensFlatt2006} type units~\cite{OwensFlatt2006}. Alice ML models components as functions of imports and static linking as an ordinary term~\cite{RossbergMissingLink2006}. MixML links mixins as a module construct~\cite{RossbergDreyerMixML2013}. \citeauthor{HirschowitzLeroy2005} compile a call-by-value CMS, a mixin calculus with mutual recursion, to a $\lambda$-calculus with records and \code{let rec}~\cite{AnconaZuccaCMS2002,HirschowitzLeroy2005}. These second-class mixins cannot be function arguments or results. Units and MixML support recursive linking, \citeauthor{OwensFlatt2006}'s units have type components, and Alice ML loads components at run time. These capabilities exceed $\sce$; none of these typed designs proves that linking compiled code agrees with the source link.

\citeauthor{FlattFelleisen1998} state no theorem. \citeauthor{OwensFlatt2006} prove type soundness, and \citeauthor{RossbergMissingLink2006} states it unproved. MixML and \citeauthor{HirschowitzLeroy2005} prove only type safety, via typed translations. CMS and MM, a call-by-value mixin calculus, have reduction semantics~\cite{AnconaZuccaCMS2002,HirschowitzLeroyWellsESOP04}, but no compiler is proved against them. \citeauthor{LeeFriedmanICFP96} give two untyped calculi, $\lambda\mathrm{C}$ and $\lambda\mathrm{CD}$~\cite{LeeFriedmanICFP96}. Source programs in $\lambda\mathrm{C}$ are contexts (terms with holes), with no linking construct. They prove that primitive linking of compiled code agrees with compiling source contexts after meta-level hole filling. $\lambda\mathrm{CD}$ adds definitions, combined with \code{seq} and \code{rec}, whose compiled forms link compiled components. For $\lambda\mathrm{CD}$, they prove no such theorem. In $\sce$, a link is a source expression, reduced by rule \ruleref{S-LinkV}, and Corollary~\ref{thm:separate-compilation} covers it, even at run time.

Units' implementation and Alice ML's semantics represent clients as functions of their imports. Units' formal model merges source text and contains no compiled code; Alice ML's static linking is only sketched. In $\sce$, this form is the link's elaboration. \citeauthor{FlattFelleisen1998} argue informally for separate compilation of closed units; Lemma~\ref{lem:context-independence} proves that a sandboxed client compiles identically in every scope. Units and MixML check a component's body under its enclosing scope, as our open functors do~\cite{FlattFelleisen1998,RossbergDreyerMixML2013}. Alice ML instead uses a custom component manager to restrict imports at run time~\cite{RossbergMissingLink2006}. A sandboxed $\sce$ client is closed by typing and, once compiled, by its box, as in $\lame$~\cite{lambdaeOOPSLA24}, wherever it is written, and Theorem~\ref{thm:link-confinement} proves its confinement. Corollary~\ref{thm:separate-compilation} proves that the core linker agrees with source linking for every well-typed link, even under binders. We know of no other typed calculus with linking as an expression that states such a theorem.

\paragraph{Modules by elaboration.}
Module systems are often defined by elaboration~\cite{DreyerCraryHarperPOPL03,FingModules2014,Rossberg1ML2018}. \citeauthor{CraryPOPL19} proves in Coq that his elaboration of ML modules is contextually correct and fully abstract~\cite{CraryPOPL19}. These systems are richer than $\sce$, with type abstraction, sharing and generativity, and full abstraction is stronger than our behavior preservation, which relates runs, not contextual equivalence. Yet none defines a linker for compiled fragments or a linking construct of its own. \citeauthor{HarperLillibridgePOPL94} model linking as functor application and claim, without proof, that separate and whole-program compilation agree~\cite{HarperLillibridgePOPL94}. Corollary~\ref{thm:static-sc} proves a semantic counterpart for simple types.

\paragraph{Dynamic linking.}
\looseness=-1 \citeauthor{ancona03calculus} combine static and dynamic linking in one calculus, separating configuration and command languages~\cite{ancona03calculus}. They leave first-class linkset values as future work. In $\sce$, links are ordinary expressions; providers and clients are first-class values. Links under binders realize \citeauthor{CardelliTypeful89}'s suggestion of imports not restricted to the top level~\cite{CardelliTypeful89}. \citeauthor{DrossopoulouLagorioEisenbach2003}, Assemblages, Acute, dynamic rebinding and TAL/Load model run-time code loading~\cite{DrossopoulouLagorioEisenbach2003,LiuSmithECOOP04,SewellAcuteJFP07,SewellRebindingJFP08,HicksWeirichCrary2000}, covering lazy resolution, re-linking and versioning, which $\sce$ does not. They prove soundness (conjectured for Acute), and \citeauthor{DrossopoulouLagorioEisenbach2003} and dynamic rebinding also prove strategies equivalent. None proves that run-time linking of compiled code agrees with source linking.

\paragraph{Verified separate compilation.}
\looseness=-1 Verified compilers treat linking outside the source language. For SepCompCert, \citeauthor{KangEtAlPOPL16} prove in Coq that linked, separately compiled C files refine the linked source, using whole-program correctness for passes that commute with linking~\cite{KangEtAlPOPL16}. CertiCoq's linker is a term template that plugs a library into an open client, and not every CertiCoq pass commutes with it~\cite{CertiCoqICFP21}. \citeauthor{BowmanAhmedPLDI18} link by substitution and prove that closure conversion commutes with it up to equivalence~\cite{BowmanAhmedPLDI18}. CakeML compiles code at run time but does not support partial compilation and linking~\cite{CakeMLEvalPLDI23}. SepCompCert and CertiCoq verify optimizing pipelines, far beyond our compiler. SepCompCert also relates I/O traces. $\sce$ has no I/O, and relates results and divergence. Still, none has a linking construct in its source, so commutation needs its own proof or fails. Nor do their theorems cover dynamic links. In $\sce$, rule \ruleref{CL-Link} wraps compiled fragments in a fixed $\lame$ template, exactly the term that \ruleref{Elab-Link} assigns to the source link. Commutation is thus a design property rather than a proof obligation, and Theorems~\ref{thm:semantics-preservation} and~\ref{thm:behavior-preservation} give correctness, static and dynamic.

\citeauthor{PattersonAhmedICFP19} distinguish linking with code from the same compiler, other compilers and other languages~\cite{PattersonAhmedICFP19}. Ours is of the first kind, whereas Compositional CompCert, Pilsner, \citeauthor{10.1007/978-3-642-54833-8_8}, and CompCertM cover the other kinds~\cite{StewartEtAlPOPL15,10.1145/2784731.2784764,10.1007/978-3-642-54833-8_8,CompCertMPOPL20}. \citeauthor{PattersonAhmedICFP19} warn that a theorem whose linker compiles source code is meaningless, as readers must trust the compiler. \ruleref{CL-Link} sees only compiled fragments and interfaces and emits a fixed $\lame$ template.

\paragraph{First-class environments.}
First-class environments turn bindings into values~\cite{GelernterJL87}; untyped languages use them for modules and run-time linking~\cite{JagannathanTOPLAS94}. \citeauthor{QueinnecDeRoure96} represent modules as functions of environments supplying their free variables~\cite{QueinnecDeRoure96}. Their \code{import} evaluates code in a supplied environment chained to the lexical one, but no theorem relates it to compiled code. The typed calculi of \citeauthor{NishizakiRIMS94} and \citeauthor{10.1007/3-540-44716-4_23} prove subject reduction and confluence~\cite{NishizakiRIMS94,10.1007/3-540-44716-4_23}, and \citeauthor{HashimotoOhoriTCS01} suggest contexts for run-time linking~\cite{HashimotoOhoriTCS01}, but none states a linking theorem. $\lame$ encodes sandboxed modules as boxed functions, with no module types and no link~\cite{TanOliveiraECOOP2023,lambdaeOOPSLA24}. Like definitional interpreters~\cite{AminRompfPOPL17}, its metatheory avoids term substitution. \citet{tan2026polymorphicfirstclassenvironments} extend first-class environments to System~F, supporting polymorphic modules with transparent type components, dynamic composition and sandboxing. Closure conversion also pairs closed code with environments, but translates a substitution-based source~\cite{MinamideMorrisettHarperPOPL96}. Over $\lame$, $\sce$ adds the module language, links, the core linker and the separate compilation corollary. Commutation itself needs no first-class environments: any compositional translation of the link would do. They determine which links a program can write. A link can take the current scope $\query$ as its provider, at a static type, and Corollary~\ref{thm:separate-compilation} covers it. A linked library can be the scope of a computation, as in \code{box lib in e}. Because a closure captures its whole environment, the box $\unit \triangleright (\lambda \elab{A}.\; e')$ empties the environment of a compiled sandboxed client, even under a binder.
 \section{Conclusion}\label{sec:conclusion}

\looseness=-1 Separate compilation correctness needs an account of linking. Linking as
evaluation makes that account the source semantics. The core linker
emits the source link's elaboration, so compilation commutes with linking.
Semantics and behavior preservation give links the same behavior at top
level, under binders or at run time. Three ingredients shape the design:
source reduction rules for linking, elaboration using only compiled
operands and their interfaces, and contexts and interfaces as ordinary types,
as in $\lame$~\cite{TanOliveiraECOOP2023,lambdaeOOPSLA24}.
Earlier work we know of has at most two.
\citeauthor{Cardelli97}'s linksets, SMLSC and verified compilers lack
source-level linking, and MixML and Alice ML define it only by translation.
Units has the construct but merges source text. The first two yield separate compilation correctness, and the third
decides which links a program can write.

Lean checks every proof, and the WebAssembly backend needs no linking
instruction. Our results cover simple types. A
concrete next step is subtyping, through an interface restriction.
Polymorphic first-class environments~\cite{tan2026polymorphicfirstclassenvironments}
provide a starting point for extending our results to polymorphic modules.
Sealing and higher-kinded polymorphism could build on a target that keeps
first-class environments and adds the type operators of System $F_\omega$, with an
elaboration in the style of F-ing modules~\cite{FingModules2014}.

\clearpage
\appendix

\section{Big-Step Semantics of \texorpdfstring{$\sce$}{lambda-SCE}}\label{app:sce-bigstep}

Figure~\ref{fig:bstep} gives the big-step semantics used in
Section~\ref{sec:separate-compilation}. The lemmas below establish
its equivalence with small-step evaluation (Theorem~\ref{thm:equiv})
and show that evaluation produces values.

\begin{figure}[!htbp]
\centering
\begin{rulefigure}
\ruleheading{\SBig{v}{e}{v'}}{Big-step evaluation}
\rulerow{
  \inferrule*[lab=\rulelabel{Eval-Val}]
    { }
    {\SBig{v}{v'}{v'}}
  \quad
  \inferrule*[lab=\rulelabel{Eval-Env}]
    { }
    {\SBig{v}{\query}{v}}
  \quad
  \inferrule*[lab=\rulelabel{Eval-Idx}]
    {\SBig{v}{e}{v_1} \\ \lookupV(v_1, n) = v_2}
    {\SBig{v}{e.n}{v_2}}
  \quad
  \inferrule*[lab=\rulelabel{Eval-Lam}]
    { }
    {\SBig{v}{\lambda A.\; e}{\clos{v}{A}{e}}}}
\rulerow{
  \inferrule*[lab=\rulelabel{Eval-App}]
    {\SBig{v}{e_1}{\clos{v_1}{A}{e}} \\
     \SBig{v}{e_2}{v_2} \\
     \SBig{v_1 \mrg v_2}{e}{v'}}
    {\SBig{v}{e_1\; e_2}{v'}}
  \qquad
  \inferrule*[lab=\rulelabel{Eval-Box}]
    {\SBig{v}{e_1}{v_1} \\ \SBig{v_1}{e_2}{v_2}}
    {\SBig{v}{e_1 \triangleright e_2}{v_2}}}
\rulerow{
  \inferrule*[lab=\rulelabel{Eval-Mrg}]
    {\SBig{v}{e_1}{v_1} \\ \SBig{v \mrg v_1}{e_2}{v_2}}
    {\SBig{v}{e_1 \mrg e_2}{v_1 \mrg v_2}}
  \qquad
  \inferrule*[lab=\rulelabel{Eval-Rcd}]
    {\SBig{v}{e}{v_1}}
    {\SBig{v}{\lrec{l}{e}}{\lrec{l}{v_1}}}
  \qquad
  \inferrule*[lab=\rulelabel{Eval-Sel}]
    {\SBig{v}{e}{v_1} \\ \selj{v_1}{l}{v_2}}
    {\SBig{v}{e.l}{v_2}}}
\rulerow{
  \inferrule*[width=0.31\linewidth,lab=\rulelabel{Eval-Let}]
    {\SBig{v}{e_1}{v_1} \\ \SBig{v \mrg v_1}{e_2}{v_2}}
    {\SBig{v}{\letb{e_1}{e_2}}{v_2}}
  \qquad
  \inferrule*[width=0.31\linewidth,lab=\rulelabel{Eval-Open}]
    {\SBig{v}{e_1}{\lrec{l}{v_1}} \\ \SBig{v \mrg v_1}{e_2}{v_2}}
    {\SBig{v}{\openm{e_1}{e_2}}{v_2}}
  \qquad
  \inferrule*[lab=\rulelabel{Eval-NMrg}]
    {\SBig{v}{e_1}{v_1} \\ \SBig{v}{e_2}{v_2}}
    {\SBig{v}{e_1 \nmrg e_2}{v_1 \mrg v_2}}}
\rulerow{
  \inferrule*[lab=\rulelabel{Eval-Str}]
    {\SBig{v}{e}{v'}}
    {\SBig{v}{\mstruct{e}}{\mstruct{v'}}}
  \qquad
  \inferrule*[lab=\rulelabel{Eval-Fun}]
    { }
    {\SBig{v}{\mfunctor{s}{A}{e}}{\mclos{\ambm{s}{v}}{A}{e}}}}
\rulerow{
  \inferrule*[width=0.45\linewidth,lab=\rulelabel{Eval-MApp}]
    {\SBig{v}{e_1}{\mclos{v_1}{A}{e}} \\
     \SBig{v}{e_2}{v_2} \\
     \SBig{v_1 \mrg v_2}{e}{v'}}
    {\SBig{v}{\mapp{e_1}{e_2}}{v'}}
  \qquad
  \inferrule*[width=0.45\linewidth,lab=\rulelabelx{Eval-Link-full}{Eval-Link}]
    {\SBig{v}{e_1}{v_1} \\
     \SBig{v}{e_2}{\mclos{v_2}{\rcd{l}{A}}{e}} \\
     \selj{v_1}{l}{v_l} \\
     \SBig{v_2 \mrg \lrec{l}{v_l}}{e}{v_3}}
    {\SBig{v}{\mlink{e_1}{e_2}}{v_1 \mrg v_3}}}
\end{rulefigure}
\caption{Big-step semantics of $\sce$.}
\Description{Big-step rules for expressions, environments, modules, and linking.}
\label{fig:bstep}
\end{figure}

\begin{lemma}[Soundness of Big-Step]\label{lem:sbig-sound}
If\/ $\SBig{v}{e}{v'}$ then $\SMStep{v}{e}{v'}$.
\end{lemma}

\begin{lemma}[Completeness of Big-Step]\label{lem:sbig-complete}
If\/ $\SMStep{v}{e}{e'}$ and $e'$ is a value, then $\SBig{v}{e}{e'}$.
\end{lemma}

\begin{lemma}[Evaluation Produces Values]\label{lem:eval-value}
If\/ $\SBig{v}{e}{e'}$ then $e'$ is a value.
\end{lemma}

\section{Extensions}\label{app:extensions}

The following rules complete the extensions of Section~\ref{sec:extensions}.

\subsection{Unions, Recursive Functions, and Recursive Types}
\label{app:ext-lang}

Unions and recursion extend both calculi with the syntax of
Figure~\ref{fig:ext-syntax} and the following runtime forms:

\syntaxrow{
\begin{array}{@{}l@{\qquad}rcl@{}}
\syntaxcat{Expressions} & e & ::= & \cdots \mid \fixclos{v}{A}{B}{e} \\
\syntaxcat{Values} & v & ::= & \cdots \mid \injl{B}{v} \mid \injr{A}{v} \mid \fixclos{v}{A}{B}{e} \mid \tfold{\mu\alpha.A}{v} \\
\syntaxcat{Frames} & F & ::= & \cdots \mid \injl{B}{\hole} \mid \injr{A}{\hole} \mid \scase{\hole}{e_1}{e_2} \mid \tfold{\mu\alpha.A}{\hole} \mid \tunfold{\hole}
\end{array}
}

In the paper, types in judgments and annotations are closed: every type
variable is bound by an enclosing $\mu$. We write $A[B/\alpha]$ for
capture-avoiding type substitution. The mechanization represents type
variables by de Bruijn indices, and its fold constructor stores the
body of the recursive-type annotation. The lemma \texttt{elab\_substTyp}
proves that type translation commutes with the mechanized substitution
at any depth.

\begin{ruleblock}
\ruleheading{\Elab{\Gamma}{e}{A}{e'}}{Elaboration: extensions}
\rulerow{
  \inferrule*[lab=\rulelabel{Elab-InjL}]
    {\Elab{\Gamma}{e}{A}{e'}}
    {\Elab{\Gamma}{\injl{B}{e}}{A \vee B}{\injl{\elab{B}}{e'}}}
  \qquad
  \inferrule*[lab=\rulelabel{Elab-InjR}]
    {\Elab{\Gamma}{e}{B}{e'}}
    {\Elab{\Gamma}{\injr{A}{e}}{A \vee B}{\injr{\elab{A}}{e'}}}}
\rulerow{
  \inferrule*[width=0.45\linewidth,lab=\rulelabel{Elab-Case}]
    {\Elab{\Gamma}{e}{A \vee B}{e'} \\
     \Elab{\Gamma \mathbin{\&} A}{e_1}{C}{e_1'} \\
     \Elab{\Gamma \mathbin{\&} B}{e_2}{C}{e_2'}}
    {\Elab{\Gamma}{\scase{e}{e_1}{e_2}}{C}{\scase{e'}{e_1'}{e_2'}}}
  \quad
  \inferrule*[lab=\rulelabel{Elab-Fix}]
    {\Elab{\Gamma \mathbin{\&} (A \to B) \mathbin{\&} A}{e}{B}{e'}}
    {\Elab{\Gamma}{\fixlam{A}{B}{e}}{A \to B}{\fixlam{\elab{A}}{\elab{B}}{e'}}}}
\rulerow{
  \inferrule*[lab=\rulelabel{Elab-FixClos}]
    {\Elab{\varepsilon}{v}{\Gamma_1}{v'} \\
     \Elab{\Gamma_1 \mathbin{\&} (A \to B) \mathbin{\&} A}{e}{B}{e'}}
    {\Elab{\Gamma}{\fixclos{v}{A}{B}{e}}{A \to B}
      {\fixclos{v'}{\elab{A}}{\elab{B}}{e'}}}}
\rulerow{
  \inferrule*[lab=\rulelabel{Elab-Fold}]
    {\Elab{\Gamma}{e}{A[\mu\alpha.A/\alpha]}{e'}}
    {\Elab{\Gamma}{\tfold{\mu\alpha.A}{e}}{\mu\alpha.A}{\tfold{\elab{\mu\alpha.A}}{e'}}}
  \qquad
  \inferrule*[lab=\rulelabel{Elab-Unfold}]
    {\Elab{\Gamma}{e}{\mu\alpha.A}{e'}}
    {\Elab{\Gamma}{\tunfold{e}}{A[\mu\alpha.A/\alpha]}{\tunfold{e'}}}}
\end{ruleblock}

The core typing rules are obtained by replacing each judgment
$\Elab{\Gamma}{e}{A}{e'}$ with $\HasType{\Gamma}{e}{A}$ and restricting
all metavariables to the core syntax. Type preservation for fold and
unfold uses the following lemma.

\begin{lemma}[Translation Commutes with Substitution]\label{lem:elab-subst}
If\/ $\mu\alpha.A$ is closed, then
\[
\elab{A[\mu\alpha.A/\alpha]} = \elab{A}\,[\elab{\mu\alpha.A}/\alpha].
\]
\end{lemma}

\begin{ruleblock}
\ruleheading{\SStep{v}{e}{e'}}{Small-step reduction: extensions}
\rulerow{
  \inferrule*[lab=\rulelabel{S-CaseL}]
    { }
    {\SStep{v}{\scase{\injl{B}{v_1}}{e_1}{e_2}}{(v \mrg v_1) \triangleright e_1}}
  \qquad
  \inferrule*[lab=\rulelabel{S-CaseR}]
    { }
    {\SStep{v}{\scase{\injr{A}{v_1}}{e_1}{e_2}}{(v \mrg v_1) \triangleright e_2}}}
\rulerow{
  \inferrule*[lab=\rulelabel{S-Fix}]
    { }
    {\SStep{v}{\fixlam{A}{B}{e}}{\fixclos{v}{A}{B}{e}}}
  \qquad
  \inferrule*[lab=\rulelabel{S-UnfoldV}]
    { }
    {\SStep{v}{\tunfold{(\tfold{\mu\alpha.A}{v_1})}}{v_1}}}
\rulerow{
  \inferrule*[lab=\rulelabel{S-FixBeta}]
    { }
    {\SStep{v}{\fixclos{v_2}{A}{B}{e}\; v_1}
      {(v_2 \mrg \fixclos{v_2}{A}{B}{e} \mrg v_1) \triangleright e}}}
\end{ruleblock}

The same frames and rules extend core reduction, with $\cvdash$
in place of $\svdash$ throughout.

\begin{ruleblock}
\ruleheading{\SBig{v}{e}{v'}}{Big-step evaluation: extensions}
\rulerow{
  \inferrule*[lab=\rulelabel{Eval-InjL}]
    {\SBig{v}{e}{v_1}}
    {\SBig{v}{\injl{B}{e}}{\injl{B}{v_1}}}
  \qquad
  \inferrule*[lab=\rulelabel{Eval-InjR}]
    {\SBig{v}{e}{v_1}}
    {\SBig{v}{\injr{A}{e}}{\injr{A}{v_1}}}}
\rulerow{
  \inferrule*[lab=\rulelabel{Eval-CaseL}]
    {\SBig{v}{e}{\injl{B}{v_1}} \\
     \SBig{v \mrg v_1}{e_1}{v'}}
    {\SBig{v}{\scase{e}{e_1}{e_2}}{v'}}
  \qquad
  \inferrule*[lab=\rulelabel{Eval-CaseR}]
    {\SBig{v}{e}{\injr{A}{v_1}} \\
     \SBig{v \mrg v_1}{e_2}{v'}}
    {\SBig{v}{\scase{e}{e_1}{e_2}}{v'}}}
\rulerow{
  \inferrule*[lab=\rulelabel{Eval-Fix}]
    { }
    {\SBig{v}{\fixlam{A}{B}{e}}{\fixclos{v}{A}{B}{e}}}
  \qquad
  \inferrule*[width=0.45\linewidth,lab=\rulelabel{Eval-FixApp}]
    {v_f = \fixclos{v_1}{A}{B}{e} \\
     \SBig{v}{e_1}{v_f} \\
     \SBig{v}{e_2}{v_2} \\
     \SBig{v_1 \mrg v_f \mrg v_2}{e}{v'}}
    {\SBig{v}{e_1\; e_2}{v'}}}
\rulerow{
  \inferrule*[lab=\rulelabel{Eval-Fold}]
    {\SBig{v}{e}{v_1}}
    {\SBig{v}{\tfold{\mu\alpha.A}{e}}{\tfold{\mu\alpha.A}{v_1}}}
  \qquad
  \inferrule*[lab=\rulelabel{Eval-Unfold}]
    {\SBig{v}{e}{\tfold{\mu\alpha.A}{v_1}}}
    {\SBig{v}{\tunfold{e}}{v_1}}}
\end{ruleblock}

Rule \ruleref{Eval-Val} covers the extended values. Replacing source
judgments with core judgments gives the corresponding big-step rules.

\subsection{Multi-Import Linking}
\label{app:ext-nary}

Import interfaces $D$ are defined in Figure~\ref{fig:ext-syntax}.

\begin{ruleblock}
\ruleheading{\linkok{I_1}{D}}{Interface satisfaction}
\rulepair{
  \inferrule*[lab=\rulelabel{OK-One}]
    {\lin{l}{A}{I_1}}
    {\linkok{I_1}{\rcd{l}{A}}}
}{
  \inferrule*[lab=\rulelabel{OK-More}]
    {\linkok{I_1}{D} \\ \lin{l}{A}{I_1}}
    {\linkok{I_1}{D \mathbin{\&} \rcd{l}{A}}}
}
\end{ruleblock}

\subsubsection*{Elaboration and Core Linking.}

Figure~\ref{fig:ext-nary} generalizes \eqref{eq:linkc}: $\wire_k(D)$
builds the import package from a provider at index $k$.
Theorem~\ref{thm:core-linkn-typed} establishes link typing under
fragment typing and interface satisfaction. Unlike \ruleref{CL-Link}, rule
\ruleref{CL-LinkN} has no premise: Theorems~\ref{thm:core-linkn-typed}
and~\ref{thm:static-sc-n} assume the satisfaction $\linkok{I_1}{D}$ instead.

\begin{figure}[!htbp]
\centering
\begin{rulefigure}
\ruleheading{\Elab{\Gamma}{e}{A}{e'},\quad\clinkn{\elab{I_1}}{e_1}{\elab{D} \to \elab{I_2}}{e_2}{e_\ell}}{Multi-import linking}
\rulerow{
  \inferrule*[lab=\rulelabel{Elab-LinkN}]
    {\Elab{\Gamma}{e_1}{I_1}{e_1'} \\
     \Elab{\Gamma}{e_2}{D \marr I_2}{e_2'} \\
     \linkok{I_1}{D}}
    {\STyp{\Gamma}{\mlinkn{e_1}{e_2}}{I_1 \mathbin{\&} I_2} \\
     \rightsquigarrow\; \cterm{
       \linkc(\elab{I_1}, \elab{D}, \elab{I_2}, e_1', e_2')}}
}
\rulerow{
  \inferrule*[lab=\rulelabel{CL-LinkN}]
    { }
    {\clinkn{\elab{I_1}}{e_1}{\elab{D} \to \elab{I_2}}{e_2}
      {\linkc(\elab{I_1}, \elab{D}, \elab{I_2}, e_1, e_2)}}
}
\syntaxrow{
\begin{array}{rcl}
\wire_{k}(\rcd{l}{A}) &\triangleq& \lrec{l}{\underline{k}.l} \\[2pt]
\wire_{k}(D \mathbin{\&} \rcd{l}{A}) &\triangleq&
  \wire_{k}(D) \mrg \lrec{l}{\underline{k{+}1}.l} \\[2pt]
\linkc(I_1, D, I_2, e_1, e_2) &\triangleq&
  (\lambda I_1.\; \lambda (D \to I_2).\;
    \underline{1} \mrg (\underline{1}\; \wire_{0}(D)))\; e_1\; e_2
\end{array}
}
\end{rulefigure}
\caption{Multi-import linking: elaboration, core linking, and the linking
term for an import interface $D$. The term of \eqref{eq:linkc} is the
instance $D = \rcd{l}{A}$.}
\Description{Source elaboration and core linking use the same term. A wire
selects the provider fields in the order required by the import interface.}
\label{fig:ext-nary}
\end{figure}

The right operand of each dependent merge uses index $k+1$, since
the left operand adds one environment entry. For
$D = \rcd{\code{x}}{\Int} \mathbin{\&} \rcd{\code{y}}{\Int}$, the wire is
$\wire_0(D) = \lrec{\code{x}}{\underline{0}.\code{x}} \mrg \lrec{\code{y}}{\underline{1}.\code{y}}$,
and the provider $\lrec{\code{y}}{2} \mrg \lrec{\code{x}}{1} \mrg \lrec{\code{z}}{3}$
supplies the package $\lrec{\code{x}}{1} \mrg \lrec{\code{y}}{2}$.
\FloatBarrier

\subsubsection*{Semantics.}
The judgment $\selpkg(v,D)=p$ states that selecting the imports of $D$
from provider $v$ gives the package $p$.

\begin{ruleblock}
\ruleheading{\selpkg(v, D) = p}{Package extraction}
\rulepair{
  \inferrule*[lab=\rulelabel{SP-One}]
    {\selj{v}{l}{v_l}}
    {\selpkg(v, \rcd{l}{A}) = \lrec{l}{v_l}}
}{
  \inferrule*[lab=\rulelabel{SP-More}]
    {\selpkg(v, D) = p \\ \selj{v}{l}{v_l}}
    {\selpkg(v, D \mathbin{\&} \rcd{l}{A}) = p \mrg \lrec{l}{v_l}}
}
\end{ruleblock}

Two additional frames evaluate the provider and then the client under
the ambient environment:
\syntaxrow{
\begin{array}{@{}l@{\qquad}rcl@{}}
\syntaxcat{Frames} & F & ::= & \cdots \mid \mlinkn{\hole}{e} \mid \mlinkn{v}{\hole}.
\end{array}
}
\begin{ruleblock}
\ruleheading{\SStep{v}{e}{e'}}{Multi-import linking}
\rulerow{
  \inferrule*[lab=\rulelabel{S-LinkNV}]
    {\selpkg(v_1, D) = p}
    {\SStep{v}{\mlinkn{v_1}{\mclos{v_2}{D}{e}}}
      {v_1 \mrg ((v_2 \mrg p) \triangleright e)}}
}
\end{ruleblock}

\begin{ruleblock}
\ruleheading{\SBig{v}{e}{v'}}{Multi-import linking}
\rulerow{
  \inferrule*[lab=\rulelabel{Eval-LinkN}]
    {\SBig{v}{e_1}{v_1} \\
     \SBig{v}{e_2}{\mclos{v_2}{D}{e}} \\
     \selpkg(v_1, D) = p \\
     \SBig{v_2 \mrg p}{e}{v_3}}
    {\SBig{v}{\mlinkn{e_1}{e_2}}{v_1 \mrg v_3}}
}
\end{ruleblock}

\subsubsection*{Properties.}

\begin{lemma}[Wire Typing]\label{lem:wire-typed}
If\/ $\linkok{I_1}{D}$ and\/ $\lookup(\Gamma, k) = \elab{I_1}$,
then\/ $\HasType{\Gamma}{\wire_k(\elab{D})}{\elab{D}}$.
\end{lemma}

\begin{theorem}[Multi-Import Link Typing]\label{thm:core-linkn-typed}
If\/ $\clinkn{\elab{I_1}}{e_1}{\elab{D} \to \elab{I_2}}{e_2}{e_\ell}$,
$\linkok{I_1}{D}$,
$\HasType{\Gamma_c}{e_1}{\elab{I_1}}$, and
$\HasType{\Gamma_c}{e_2}{\elab{D} \to \elab{I_2}}$,
then $\HasType{\Gamma_c}{e_\ell}{\elab{I_1} \mathbin{\&} \elab{I_2}}$.
\end{theorem}

\begin{theorem}[Multi-Import Static Separate Compilation]\label{thm:static-sc-n}
Let $e_2 = \mfunctor{\sandbox}{D}{e}$ with
$\Elab{\varepsilon}{e_2}{D \marr I_2}{e_2'}$.
Suppose\/ $\Elab{\varepsilon}{e_1}{I_1}{e_1'}$,
$\linkok{I_1}{D}$, and
$\clinkn{\elab{I_1}}{e_1'}{\elab{D} \to \elab{I_2}}{e_2'}{e_\ell}$.
Then clauses (a)--(c) of Corollary~\ref{thm:separate-compilation} hold for
$\mlinkn{e_1}{e_2}$ and $\cterm{e_\ell}$, with $v = u = \unit$.
\end{theorem}

\begin{proof}[Proof Sketch of Theorem~\ref{thm:static-sc-n}]
Rules \ruleref{Elab-LinkN} and \ruleref{CL-LinkN} produce the same
core term under the theorem's premises.
Theorems~\ref{thm:semantics-preservation} and~\ref{thm:behavior-preservation}
then give clauses (a)--(c).
\end{proof}

As in Corollary~\ref{thm:separate-compilation}, the result generalizes
to a provider elaborated under $\Gamma$ and evaluation under a
corresponding environment. Lemma~\ref{lem:context-independence}
reuses the client's elaboration under $\varepsilon$ unchanged.

For multi-import linking, the mechanized replacement law requires
the replacement provider to preserve every extractable package,
over all import interfaces.

\begin{samepage}
\begin{theorem}[Multi-Import Interface Replacement]\label{thm:interface-replacement-n}
Suppose $\SBig{v}{\mlinkn{e_1}{e}}{p_1 \mrg w}$,
$\SBig{v}{e_2}{p_2}$, and, for every $D$ and $q$,
$\selpkg(p_1,D)=q$ implies $\selpkg(p_2,D)=q$. Then
\[
\SBig{v}{\mlinkn{e_2}{e}}{p_2 \mrg w}.
\]
If also $\STyp{\Gamma}{\mlinkn{e_2}{e}}{A}$,
$\STyp{\varepsilon}{v}{\Gamma}$, and
$\SBig{v}{\mlinkn{e_2}{e}}{p_2 \mrg w'}$, then $w=w'$.
\end{theorem}

\begin{proof}[Proof Sketch]
Reuse the client and body evaluations in rule \ruleref{Eval-LinkN},
with the replacement provider and the same import package.
The equality follows from determinism (Theorem~\ref{thm:determinism}).
\end{proof}
\end{samepage}

The first part of Theorem~\ref{thm:interface-replacement-n} needs no
typing, and the theorem concerns only terminating runs. For typed
provider values, a two-way law holds, the multi-import counterpart of
Theorem~\ref{thm:interface-abstraction}. Providers that supply the same
import package give the client the same exports, whatever they supply at
other interfaces, and one link diverges exactly when the other does.

\begin{theorem}[Multi-Import Interface Abstraction]\label{thm:interface-abstraction-n}
Let $p_1$ and $p_2$ be values. Suppose
$\STyp{\Gamma}{e}{D \marr B}$,
$\STyp{\varepsilon}{v}{\Gamma}$, and, for $i = 1,2$,
$\STyp{\varepsilon}{p_i}{P_i}$, $\linkok{P_i}{D}$, and
$\selpkg(p_i,D)=q$. Then, for every value $w$,
\[
\SBig{v}{\mlinkn{p_1}{e}}{p_1 \mrg w}
\quad\Longleftrightarrow\quad
\SBig{v}{\mlinkn{p_2}{e}}{p_2 \mrg w}
\]
and $\SDiv{v}{\mlinkn{p_1}{e}}$ if and only if $\SDiv{v}{\mlinkn{p_2}{e}}$.
\end{theorem}

\section{Implementation Details}\label{app:implementation}

We give the unit-linking construction and WasmGC compilation used by
\textsf{SCE-Lang} (Section~\ref{sec:implementation}).

\subsection{Units and Linking}\label{app:units-linking}
An importing file is represented by the sandboxed functor
$\mfunctor{\sandbox}{\rcd{M_1}{A_1} \mathbin{\&} \cdots \mathbin{\&}
\rcd{M_k}{A_k}}{e}$, with each $A_i$ read from $M_i$'s interface file.
An import-free file is a structure boxed under the empty environment.

Let $e_1, \ldots, e_n$ be units in dependency order, with $n\geq 1$
and $e_1$ import-free. Each $e_i$ elaborates independently under
$\varepsilon$ to $\cterm{c_i}$, at type $E_i$ if import-free and
$D_i\marr E_i$ otherwise. Write $S_i$, $\cterm{L_i}$, and $I_i$ for
the source composition, core composition, and accumulated export type.
Initially,
\[
S_1 = e_1
\qquad
\cterm{L_1} = \cterm{c_1}
\qquad
I_1 = E_1.
\]
For an importing unit $e_{i+1}$, require $\linkok{I_i}{D_{i+1}}$ and set
\[
\begin{aligned}
S_{i+1} &= \mlinkn{S_i}{e_{i+1}}, \\
\cterm{L_{i+1}} &=
  \cterm{\linkc(\elab{I_i},\, \elab{D_{i+1}},\, \elab{E_{i+1}},\,
    L_i,\, c_{i+1})}.
\end{aligned}
\]
For an import-free unit, set $S_{i+1} = S_i \nmrg e_{i+1}$ and
construct $\cterm{L_{i+1}}$ from $\cterm{L_i}$ and $\cterm{c_{i+1}}$
using the encoding of rule \ruleref{Elab-NMrg}. In either case,
$I_{i+1} = I_i \mathbin{\&} E_{i+1}$.
For binary chains, each importing unit has a singleton interface
$D_{i+1} = \rcd{l}{A}$; require $\lin{l}{A}{I_i}$ and use
$S_{i+1} = \mlink{S_i}{e_{i+1}}$. The core construction is unchanged.

\begin{theorem}[Unit-Chain Correctness]\label{thm:unit-chain}
For a binary chain constructed above,
$\Elab{\varepsilon}{S_n}{I_n}{L_n}$, and clauses (a)--(c) of
Corollary~\ref{cor:whole-program} hold for $S_n$ and $L_n$ at type $I_n$.
\end{theorem}

\begin{theorem}[Multi-Import Unit-Chain Correctness]\label{thm:unit-chain-n}
For a multi-import chain constructed above,
$\Elab{\varepsilon}{S_n}{I_n}{L_n}$, and clauses (a)--(c) of
Corollary~\ref{cor:whole-program} hold for $S_n$ and $L_n$ at type $I_n$.
\end{theorem}

\begin{proof}[Proof Sketch]
Induct on the chain, starting from the first unit's elaboration.
Each extension uses \ruleref{Elab-Link} or \ruleref{Elab-LinkN} for an
importing unit, and \ruleref{Elab-NMrg} for an import-free unit.
Corollary~\ref{cor:whole-program} then gives clauses (a)--(c).
\end{proof}

For WebAssembly, a link module imports unit entry
points and compiles the same composition with environment projections
for their results (Appendix~\ref{app:example-wasm}).

\subsection{Compilation to WebAssembly}\label{app:wasm}
Figure~\ref{fig:wasm-full} compiles well-typed program terms of extended
$\lame$ (Appendix~\ref{app:extensions}); runtime closures are not
compilation inputs. Field selection uses the inferred type of its operand.
The Wasm backend represents integers and performs arithmetic with
\code{i32}, so the literal rule is restricted to signed 32-bit integers.

\begin{figure}[!htbp]
\centering
\footnotesize
\setlength{\wasmwidth}{0.68\linewidth}
\begin{tabular}{@{}p{0.26\linewidth}@{\;\;$=$\;\;}l@{}}
$\comp{\query}{\rho}$ &
  \wasmblock{\wi{local.get} $\rho$} \\[4pt]
$\comp{\unit}{\rho}$ &
  \wasmblock{\wi{global.get} \wty{unit}} \\[4pt]
$\comp{i}{\rho}$ &
  \wasmblock{\wi{i32.const} $\tau_{\code{int}}$ \ws
    \wi{i32.const} $i$ \ws \wi{struct.new} \wty{I32Box}} \\[4pt]
$\comp{e.n}{\rho}$ &
  \wasmblock{$\comp{e}{\rho}$ \ws $\mathcal{L}^{n}$ \ws $\mathcal{R}$} \\[4pt]
$\comp{e.l}{\rho}$ &
  \wasmblock{$\comp{e}{\rho}$ \ws $\mathcal{P}(A,l)$ \ws
    \wi{ref.cast} (\wi{ref} \wty{Lrec}) \\
    \wi{struct.get} \wty{Lrec} \wfield{value}} \\[4pt]
$\comp{\lrec{l}{e}}{\rho}$ &
  \wasmblock{\wi{i32.const} $\tau_{\code{lrec}}$ \ws
    $\mathcal{B}(l)$ \ws $\comp{e}{\rho}$ \ws
    \wi{struct.new} \wty{Lrec}} \\[4pt]
$\comp{e_1 \triangleright e_2}{\rho}$ &
  \wasmblock{$\comp{e_1}{\rho}$ \ws \wi{local.set} $\rho'$ \ws
    $\comp{e_2}{\rho'}$} \\[4pt]
$\comp{e_1 \mrg e_2}{\rho}$ &
  \wasmblock{$\comp{e_1}{\rho}$ \ws \wi{local.set} $x$ \\
    \wi{i32.const} $\tau_{\code{mrg}}$ \ws
    \wi{local.get} $\rho$ \ws \wi{local.get} $x$ \\
    \wi{struct.new} \wty{Pair} \ws \wi{local.set} $\rho'$ \\
    \wi{i32.const} $\tau_{\code{mrg}}$ \ws
    \wi{local.get} $x$ \ws $\comp{e_2}{\rho'}$ \ws
    \wi{struct.new} \wty{Pair}} \\[4pt]
$\comp{\lambda A.\;e}{\rho}$ &
  \wasmblock{\wi{i32.const} $\tau_{\code{clos}}$ \ws
    \wi{ref.func} \wty{f} \ws \wi{local.get} $\rho$ \ws
    \wi{struct.new} \wty{Clos} \\
    \textrm{where the fresh lifted function} \wty{f}$(\code{self},x)$ \textrm{is} \\
    \quad \wi{i32.const} $\tau_{\code{mrg}}$ \ws
      \wi{local.get} \code{self} \ws \wi{ref.cast} (\wi{ref} \wty{Clos}) \\
    \quad \wi{struct.get} \wty{Clos} \wfield{env} \ws
      \wi{local.get} $x$ \ws \wi{struct.new} \wty{Pair} \\
    \quad \wi{local.set} $\rho'$ \ws $\comp{e}{\rho'}$} \\[4pt]
$\comp{e_1\;e_2}{\rho}$ &
  \wasmblock{$\comp{e_1}{\rho}$ \ws \wi{local.tee} $c$ \ws
    $\comp{e_2}{\rho}$ \ws \wi{local.get} $c$ \\
    \wi{ref.cast} (\wi{ref} \wty{Clos}) \ws
    \wi{struct.get} \wty{Clos} \wfield{fn} \ws
    \wi{call\_ref} \wty{Fn}}
\end{tabular}

\medskip
\begin{tabular}{@{}l@{\;\;$=$\;\;}l@{}}
$\mathcal{L}$ & \wi{ref.cast} (\wi{ref} \wty{Pair}) \ws
  \wi{struct.get} \wty{Pair} \wfield{left} \\
$\mathcal{R}$ & \wi{ref.cast} (\wi{ref} \wty{Pair}) \ws
  \wi{struct.get} \wty{Pair} \wfield{right} \\
$\mathcal{B}(l)$ & \wi{i32.const} $\mathit{off}_l$ \ws
  \wi{i32.const} $\mathit{len}_l$ \ws
  \wi{array.new\_data} \wty{Bytes} \wty{strings}
\end{tabular}
\[
\begin{array}{@{}rcll@{}}
\mathcal{P}(\rcd{l}{A},l) &=& [\,] \\
\mathcal{P}(A \mathbin{\&} B,l) &=& \mathcal{L}\;\mathcal{P}(A,l)
  & \text{if }l\in\labels{A},\ l\notin\labels{B} \\
\mathcal{P}(A \mathbin{\&} B,l) &=& \mathcal{R}\;\mathcal{P}(B,l)
  & \text{if }l\notin\labels{A},\ l\in\labels{B}.
\end{array}
\]
\caption{WasmGC compilation of extended $\lame$ (1/2): the base calculus.
The local $\rho$ holds the environment, and $\tau_k$ is the tag for
constructor $k$. Introduced locals are fresh; each compiled term leaves
one value reference on the stack. In selection, $A$ is the inferred type of $e$.
Juxtaposition concatenates instruction sequences; $\mathcal{L}^{n}$
repeats $\mathcal{L}$ $n$ times. The pair
$(\mathit{off}_l,\mathit{len}_l)$ locates $l$ in the passive data segment
\wty{strings}.}
\Description{WasmGC instruction sequences for environments, integers,
records, boxes, dependent merges, functions, and application.}
\label{fig:wasm-full}
\end{figure}

\begin{figure}[!htbp]
\ContinuedFloat
\centering
\footnotesize
\setlength{\wasmwidth}{0.68\linewidth}
\begin{tabular}{@{}p{0.26\linewidth}@{\;\;$=$\;\;}l@{}}
$\comp{\fixlam{A}{B}{e}}{\rho}$ &
  \wasmblock{\wi{i32.const} $\tau_{\code{fclos}}$ \ws
    \wi{ref.func} \wty{f} \ws \wi{local.get} $\rho$ \\
    \wi{struct.new} \wty{Clos} \\
    \textrm{where the fresh lifted function} \wty{f}$(\code{self},x)$ \textrm{is} \\
    \quad \wi{i32.const} $\tau_{\code{mrg}}$ \ws
      \wi{i32.const} $\tau_{\code{mrg}}$ \\
    \quad \wi{local.get} \code{self} \ws \wi{ref.cast} (\wi{ref} \wty{Clos}) \ws
      \wi{struct.get} \wty{Clos} \wfield{env} \\
    \quad \wi{local.get} \code{self} \ws
      \wi{struct.new} \wty{Pair} \ws \wi{local.get} $x$ \\
    \quad \wi{struct.new} \wty{Pair} \ws
      \wi{local.set} $\rho'$ \ws $\comp{e}{\rho'}$} \\[5pt]
$\comp{\injl{B}{e}}{\rho}$ &
  \wasmblock{\wi{i32.const} $\tau_{\code{inl}}$ \ws
    $\comp{e}{\rho}$ \ws \wi{struct.new} \wty{Wrap}} \\[5pt]
$\comp{\injr{A}{e}}{\rho}$ &
  \wasmblock{\wi{i32.const} $\tau_{\code{inr}}$ \ws
    $\comp{e}{\rho}$ \ws \wi{struct.new} \wty{Wrap}} \\[5pt]
$\comp{\scase{e}{e_1}{e_2}}{\rho}$ &
  \wasmblock{$\comp{e}{\rho}$ \ws \wi{local.set} $u$ \\
    \wi{local.get} $u$ \ws \wi{struct.get} \wty{Val} \wfield{tag} \\
    \wi{i32.const} $\tau_{\code{inl}}$ \ws \wi{i32.eq} \\
    \wi{if} (\wi{result} (\wi{ref null} \wty{Val})) \\
    \quad $\mathcal{E}(\rho,u,\rho_1)$ \ws $\comp{e_1}{\rho_1}$ \\
    \wi{else} \\
    \quad $\mathcal{E}(\rho,u,\rho_2)$ \ws $\comp{e_2}{\rho_2}$ \\
    \wi{end}} \\[5pt]
$\comp{\tfold{\mu\alpha.A}{e}}{\rho}$ &
  \wasmblock{\wi{i32.const} $\tau_{\code{fold}}$ \ws
    $\comp{e}{\rho}$ \ws \wi{struct.new} \wty{Wrap}} \\[5pt]
$\comp{\tunfold{e}}{\rho}$ &
  \wasmblock{$\comp{e}{\rho}$ \ws \wi{ref.cast} (\wi{ref} \wty{Wrap}) \ws
    \wi{struct.get} \wty{Wrap} \wfield{value}} \\[9pt]
$\mathcal{E}(\rho,u,\rho')$ &
  \wasmblock{\wi{i32.const} $\tau_{\code{mrg}}$ \ws
    \wi{local.get} $\rho$ \ws \wi{local.get} $u$ \\
    \wi{ref.cast} (\wi{ref} \wty{Wrap}) \ws
    \wi{struct.get} \wty{Wrap} \wfield{value} \\
    \wi{struct.new} \wty{Pair} \ws \wi{local.set} $\rho'$}
\end{tabular}

\medskip
\begin{tabular}{@{}ll@{\qquad}ll@{}}
\wty{Val} & \wfield{tag}:\wi{i32}
  & \wty{I32Box} & \wfield{tag}:\wi{i32}, \wfield{n}:\wi{i32} \\
\wty{Pair} & \wfield{tag}:\wi{i32}, \wfield{left}:$V$, \wfield{right}:$V$
  & \wty{Lrec} & \wfield{tag}:\wi{i32}, \wfield{label}:$B$, \wfield{value}:$V$ \\
\wty{Clos} & \wfield{tag}:\wi{i32}, \wfield{fn}:$F$, \wfield{env}:$V$
  & \wty{Wrap} & \wfield{tag}:\wi{i32}, \wfield{value}:$V$
\end{tabular}
\[
\begin{gathered}
V = (\wi{ref null}\;\wty{Val}),\qquad
B = (\wi{ref null}\;\wty{Bytes}),\qquad
F = (\wi{ref null}\;\wty{Fn}) \\
\wty{Bytes} = (\wi{array}\;(\wi{mut}\;\wi{i8})),\qquad
\wty{Fn} : (V,V) \to V,\qquad \wty{UnitFn} : () \to V.
\end{gathered}
\]
\caption[]{WasmGC compilation of extended $\lame$ (2/2): unions and
recursion, with runtime layouts in field-index order. Struct fields are
immutable; each struct except \wty{Val} is a final subtype of \wty{Val}.
Value locals and results have type $V$. The global \wty{unit} is a
\wty{Val} with tag $3$. Tags for integers, merges, records, closures,
recursive closures, left and right injections, and folds are
$0,4,5,6,7,8,9,10$, respectively.}
\Description{WasmGC instruction sequences for recursive functions, unions,
and recursive types, followed by the runtime types and their fields.}
\end{figure}
 
\clearpage
\section{The Double Example Through Compilation}
\label{app:linking-example}

The files of Figure~\ref{fig:sepcomp} appear below in the paper's surface notation.

\par\noindent\begin{minipage}{\linewidth}
\begin{minipage}[t]{0.47\linewidth}
\srcfile{Add.scei}
\begin{srcnow}{sceiground}{appendix-e-interfaces}
\begin{lstlisting}[style=scei]
sig
  op : Int -> Int -> Int
  unit : Int
end
\end{lstlisting}
\end{srcnow}
\end{minipage}\hfill
\begin{minipage}[t]{0.47\linewidth}
\srcfile{Double.scei}
\begin{srcnow}{sceiground}{appendix-e-interfaces}
\begin{lstlisting}[style=scei]
sig double : Int -> Int end
\end{lstlisting}
\end{srcnow}
\end{minipage}

\smallskip
\begin{minipage}[t]{0.47\linewidth}
\srcfile{Add.sce}
\begin{srcnow}{sceground}{appendix-e-sources}
\begin{lstlisting}[style=sce]
module Add = struct
  let op (x y : Int) : Int = x + y
  let unit : Int = 0
end
\end{lstlisting}
\end{srcnow}
\end{minipage}\hfill
\begin{minipage}[t]{0.47\linewidth}
\srcfile{Double.sce}
\begin{srcnow}{sceground}{appendix-e-sources}
\begin{lstlisting}[style=sce]
import Add
module Double = struct
  let double (x : Int) : Int = Add.op x x
end
\end{lstlisting}
\end{srcnow}
\end{minipage}
\end{minipage}\par

\subsection{Link, Then Compile}\label{app:example-source}

The unit wrappers give the following top-level link $S$:
\par\noindent\begin{minipage}{\linewidth}
\begin{lstlisting}
link (sandbox struct
  module Add = struct
    let op (x y : Int) : Int = x + y
    let unit : Int = 0
  end
end) with sandbox functor
  (M : { Add : sig op : Int -> Int -> Int; unit : Int end }) ->
    open M in struct
      module Double = struct
        let double (x : Int) : Int = Add.op x x
      end
    end
\end{lstlisting}
\end{minipage}\par

We abbreviate the core terms below, with $\underline{n}=\query.n$.
Identity applications and the internal label \code{\%open} are retained
from the compiler output.
{\compactdisplayskips\[
\begin{aligned}
O &= \Int\to\Int\to\Int,
& \code{P} &= \rcd{\code{op}}{O}\mathbin{\&}\rcd{\code{unit}}{\Int}, \\
\code{T} &= \rcd{\code{double}}{\Int\to\Int},
& I &= \rcd{\code{Add}}{\code{P}},
\qquad J = \rcd{\code{Double}}{\code{T}}.
\end{aligned}
\]}
{\compactdisplayskips\[
\begin{aligned}
\mathit{id}_A &= \lambda A.\;\underline{0}, \\
a_+ &= \lambda\Int.\;\lambda\Int.\;\underline{1}+\underline{0}, \\
b &= \lambda\Int.\;(\underline{1}.\code{Add}.\code{op})\;
  \underline{0}\;\underline{0}, \\
r_A &= \lrec{\code{op}}{\mathit{id}_O\;a_+}
  \mrg \lrec{\code{unit}}{\mathit{id}_{\Int}\;0}, \\
r_D &= \lrec{\code{Double}}{
  \lrec{\code{double}}{\mathit{id}_{\Int\to\Int}\;b}}, \\
a &= \unit\triangleright\lrec{\code{Add}}{r_A}, \\
d &= \unit\triangleright\lambda I.\;
  (\lambda I.\;r_D)\;(\lrec{\code{\%open}}{\underline{0}}.\code{\%open}).
\end{aligned}
\]}
Compiling $S$ gives
{\compactdisplayskips\[
L =
\underbrace{\bigl(\lambda I.\;\lambda(I\to J).\;
  \underline{1}\mrg
  (\underline{1}\;\lrec{\code{Add}}{\underline{0}.\code{Add}})\bigr)}_{K}
\;a\;d.
\]}
The linked library exports \code{Add} and \code{Double}, and
$S.\code{Double}.\code{double}\;5$ evaluates to $10$.

\subsection{Compile, Then Link in the Core}\label{app:example-core}

Compiling the files independently produces:
{\compactdisplayskips\[
\begin{array}{@{}lcl@{}}
\code{Add.sceo} &\longmapsto& a : I \\
\code{Double.sceo} &\longmapsto& d : I\to J.
\end{array}
\]}
The core linker selects \code{Add} for the client's import and produces
{\compactdisplayskips\[
\begin{aligned}
L_{\mathrm{core}}
  &= \bigl(\lambda I.\;\lambda(I\to J).\;
    \underline{1}\mrg
    (\underline{1}\;\lrec{\code{Add}}{\underline{0}.\code{Add}})\bigr)
    \;a\;d \\
  &= L.
\end{aligned}
\]}
Both routes produce the same core term, with
{\compactdisplayskips\[
\EBig{\unit}{L_{\mathrm{core}}.\code{Double}.\code{double}\;5}{10}.
\]}

\subsection{Compile, Then Link in WebAssembly}\label{app:example-wasm}

The core units compile to separate modules, each exporting
\code{main : () -> (ref null \$Val)}:
{\compactdisplayskips\[
\begin{array}{@{}rcl@{\qquad\qquad}rcl@{}}
a &\longmapsto& \code{Add.wasm}
  & d &\longmapsto& \code{Double.wasm}.
\end{array}
\]}
If their entry points return $p$ and $f$, the link module constructs
$\rho=(\unit\mrg p)\mrg f$ and executes the compiled term
$K\;\underline{1}\;\underline{0}$.
The listing uses symbolic type and function names and omits shared
type declarations and rendering support.

\noindent\srcfile{Lib.wasm}
\begin{wasmsource}
\begin{lstlisting}[style=wasmcode]
(module
  ;; Shared types and rendering support omitted.
  (import "u0" "main" (func $Add.main (type $UnitFn)))
  (import "u1" "main" (func $Double.main (type $UnitFn)))
  (global $unit (ref null $Val) (struct.new $Val (i32.const 3)))
  (data $labels "Add")
  (elem declare func $link0 $link1)

  (func $main (export "main") (type $UnitFn)
    (local (ref null $Val) (ref null $Val) (ref null $Val))
    global.get $unit  local.set 0
    i32.const 4  local.get 0  call $Add.main  struct.new $Pair  local.set 0
    i32.const 4  local.get 0  call $Double.main  struct.new $Pair  local.set 0
    i32.const 6  ref.func $link0  local.get 0  struct.new $Clos  local.tee 1
    local.get 0  ref.cast (ref $Pair)  struct.get $Pair 1  ref.cast (ref $Pair)
    struct.get $Pair 2  local.get 1  ref.cast (ref $Clos)  struct.get $Clos 1
    call_ref $Fn
    local.tee 2
    local.get 0  ref.cast (ref $Pair)  struct.get $Pair 2  local.get 2
    ref.cast (ref $Clos)  struct.get $Clos 1  call_ref $Fn
  )

\end{lstlisting}
\end{wasmsource}
\par\noindent\begin{minipage}{\linewidth}
\srcfile{Lib.wasm (continued)}
\begin{wasmsource}
\begin{lstlisting}[style=wasmcode]
  (func $link0 (type $Fn)
    (local (ref null $Val))
    i32.const 4  local.get 0  ref.cast (ref $Clos)  struct.get $Clos 2
    local.get 1  struct.new $Pair  local.set 2
    i32.const 6  ref.func $link1  local.get 2  struct.new $Clos
  )

  (func $link1 (type $Fn)
    (local (ref null $Val) (ref null $Val) (ref null $Val))
    (local (ref null $Val))
    i32.const 4  local.get 0  ref.cast (ref $Clos)  struct.get $Clos 2
    local.get 1  struct.new $Pair  local.set 2
    local.get 2  ref.cast (ref $Pair)  struct.get $Pair 1  ref.cast (ref $Pair)
    struct.get $Pair 2  local.set 3
    i32.const 4  local.get 2  local.get 3  struct.new $Pair  local.set 4
    i32.const 4  local.get 3  local.get 4  ref.cast (ref $Pair)
    struct.get $Pair 1  ref.cast (ref $Pair)  struct.get $Pair 2  local.tee 5
    i32.const 5  i32.const 0  i32.const 3  array.new_data $Bytes $labels
    local.get 4  ref.cast (ref $Pair)  struct.get $Pair 2  ref.cast (ref $Lrec)
    struct.get $Lrec 2  struct.new $Lrec  local.get 5  ref.cast (ref $Clos)
    struct.get $Clos 1  call_ref $Fn
    struct.new $Pair
  )
  ;; sce.units: Add, Double
)
\end{lstlisting}
\end{wasmsource}
\end{minipage}\par
\medskip

\noindent Each unit and the link module declares the same recursive group
of ABI types (Figure~\ref{fig:wasm-full}), allowing values to pass between
instances. The host supplies the unit entry points and calls the linker:
\begin{lstlisting}[style=hostcode]
const imports = {
  u0: { main: add.exports.main },
  u1: { main: double.exports.main }
};
const lib = new WebAssembly.Instance(linkModule, imports);
const result = lib.exports.main();
\end{lstlisting}
The linked result, rendered by the implementation, is
\begin{lstlisting}[language={}]
{ Add = { op = <fun>, unit = 0 }, Double = { double = <fun> } }
\end{lstlisting}
Calling its \code{Double.double} function on $5$ returns $10$,
as in Sections~\ref{app:example-source} and~\ref{app:example-core}.
  
\end{document}